\documentclass[onecolumn,authoryear]{els-mrw}

\def\aj{AJ}                      
\def\araa{ARA\&A}       
\def\apj{ApJ}                 
\def\apjl{ApJ}                
\def\apjs{ApJS}             
\def\aap{A\&A}              
\def\mnras{MNRAS}     
\def\nat{Nature}
\def\pasp{PASP}

\usepackage{amsmath,amssymb,amsfonts,amsthm,makeidx,graphicx}
\usepackage{txfonts}
\usepackage{helvet}

\usepackage[dvipsnames]{xcolor}

\usepackage{tocloft}
\renewcommand{\contentsname}{}
\usepackage{tikz}
\usepackage{overpic}

\usepackage[colorlinks]{hyperref}

\usepackage{xcolor}
\definecolor{dark-red}{rgb}{0.9,0.0,0.0}
\definecolor{dark-blue}{rgb}{0.15,0.15,0.9}
\definecolor{dark-green}{rgb}{0.15,0.8,0.15}
\definecolor{medium-blue}{rgb}{0,0,0.9}
\hypersetup{linkcolor={dark-blue},citecolor={dark-blue}, urlcolor={medium-blue}
}

\usepackage{pgfplots}
\usepackage{pgf-spectra}

\definecolor{color1}{RGB}{202,0,32}
\definecolor{color2}{RGB}{244,165,130}
\definecolor{color3}{RGB}{146,197,222}
\definecolor{color4}{RGB}{5,113,176}

\begin{document}

\chapter*{Radial velocity technique} 

\author[1,2]{Trifon Trifonov}%
%

\address[1]{\orgname{Universt\"at Heidelberg}, \orgdiv{Landessternwarte, Zentrum f\"ur Astronomie}, \orgaddress{K\"onigstuhl 12, 69117 Heidelberg, Germany}}
\address[2]{\orgname{Sofia University ``St. Kliment Ohridski''}, \orgdiv{Department of Astronomy, Faculty of Physics}, \orgaddress{James Bourchier Blvd., BG-1164 Sofia, Bulgaria}}


\maketitle

\begin{abstract}[Abstract]
The precise Doppler method for measuring stellar radial velocities (RV) is a fundamental technique in modern astronomy. This method records a star's spectrum and
detects periodic Doppler shifts in its spectral features, which indicate the gravitational influences induced by orbiting companions. The Doppler technique has
yielded remarkable successes in exoplanet detection, uncovering a diverse array of planetary systems ranging from hot Jupiters to Neptune-mass planets and super-Earths. Having led to the discovery of over 1,100 exoplanets, the RV method is the most effective approach for measuring orbital geometries
and minimum masses, which are of fundamental importance when accessing planet formation and evolution scenarios.
However, exoplanet detection via precise RV measurements poses significant challenges, including managing various sources of interference, such as instrumental errors, and mitigating spurious Doppler shifts induced by phenomena like stellar activity. Key to this technique's advancement is instrumental calibration methods, notably precise calibration methods and ultra-stable spectrographs.
This technique holds promise in systematically exploring the domain of Jovian analogs, rocky and icy planets within the habitable zones of their parent stars, and providing crucial follow-up observations for transiting candidates detected by space missions.
The synergy between transit and Doppler measurements of exoplanets, when feasible, has
provided a comprehensive set of orbital and physical parameters for exoplanets, such as the
dynamical mass and mean planet density, which is of paramount importance in thereby
enhancing our understanding of their internal composition.
Additionally, ongoing efforts aim to improve the RV technique further by developing more
stable calibration techniques aimed at detecting Earth-like analogs around Solar-type stars that require cm\,s$^{-1}$ RV precision.
\end{abstract}

\begin{keywords}
Radial Velocity, Doppler shift, Exoplanet detection methods, Exoplanets, Spectroscopy, Orbital elements, Keplerian orbit
\end{keywords}


\noindent \rule{\textwidth}{1.0pt} \\
\noindent {\Large \bf Contents}
\hfill \break
\vspace{-3.3cm}
\tableofcontents
\vspace{0.25cm}
\noindent \rule{\textwidth}{1.0pt}

\newpage

\begin{glossary}[Learning Objectives]
This chapter aims to provide the very basic principles of the precise Doppler spectroscopy method for the detection of exoplanets around stars. By the end of this chapter, you will learn:
\begin{itemize}

 \item History of the Doppler method in exoplanet Astronomy.

 \item Impact of the precise RV method of discovering exoplanets.

 \item Basic principles of measuring precise RVs using \'{E}chelle spectrographs.

 \item How to interpret and model RV data, and derive orbital parameters.

 \item Limitations on RVs measurements due to instrumental effects and stellar activity.

 \item The future role of the RV method in modern astronomy and astrophysics.

\end{itemize}

\end{glossary}

\begin{glossary}[Glossary]

\term{Argument of periastron (or pericenter)} in the context of exoplanets $\omega$ is the angle from the ascending node of the exoplanet's orbit to its closest point to the host star, measured in the direction of motion.

\term{\'{E}chelle spectrograph} A type of modern high-resolution spectrograph that uses a second diffraction element (grating or a prism) to disperse light into multiple orders, allowing for the simultaneous observation of a wide spectral range across many narrow bandwidths.

\term{Eccentric anomaly} in the context of exoplanets is an angular parameter that represents the position of an exoplanet along its elliptical orbit, measured from the centre of the ellipse to the projection of the planet's position onto the circumscribed circle.

\term{Fabry–Pérot etalon} is used for correcting for instrumental drifts and potentially for precise wavelength calibration in \'{E}chelle spectrographs. F-P etalons can achieve RV precision of a few cm\,s$^{-1}$ when actively controlled using atomic references.

\term{Habitable zone} is an orbital zone, which allows for temperatures just right for liquid water to exist on a planet's surface.

\term{Inclination} in the context of exoplanets, the inclination angle $i$ refers to the angle between the plane of the planet's orbit around its host star and the line of sight from an observer on Earth.

\term{Iodine cell} is the best-known example of a gas absorption cell filled with iodine gas heated to about 50 C$^{\circ}$ and placed in the optical path of a spectrograph. The superimposed spectra of I$_2$ absorption lines onto the stellar spectra provide an accurate wavelength reference for precise RV measurements.

\term{Laser Frequency Comb calibrator} provides a dense array of equally spaced laser peaks over a broad bandwidth for precision and stability for wavelength calibration in \'{E}chelle spectrographs. This technology promises RV precision down to 1 cm\,s$^{-1}$.

\term{Markov chain Monte Carlo} is a computational algorithm that samples fitting parameters from a probability distribution used for Bayesian inference statistical analysis and is widely used in Exoplanet characterization to derive orbital parameter posterior distribution and their uncertainty estimates.

\term{Mean anomaly} is an angle that describes the position of a planet as a function of time, assuming the planet is moving in a circular orbit with a constant angular speed and the same period as the planet in an eccentric orbit. The mean anomaly is used to calculate the instantaneous position of an exoplanet in its eccentric orbit.

\term{SB1 \& SB2 binary stars} are spectroscopic binary star systems. In a single-lined spectroscopic binary (SB1), the primary star is much more luminous than the secondary, so only the spectral lines of the bright component are recorded. In contrast, a double-lined spectroscopic binary (SB2) has visible spectral lines from both stars, enabling the determination of the RVs of both components.

\term{Spectroscopic orbital elements} in the context of RV-derived Keplerian orbits 
the set of derivable exoplanet parameters includes the orbital period $P$, the eccentricity $e$, the argument of periastron $\omega$, the semi-amplitude $K$, and the time of periastron passage $t_p$, which together describe the shape and orientation of the exoplanet's orbit along the line of sight as well the planet's minimum mass if the stellar mass is known.

\term{Thorium-Argon lamp}  is the most commonly used example of a hollow-cathode lamp, a standard calibration source lamp used in spectroscopy. It emits a well-understood spectrum of Thorium and Argon emission lines, providing accurate wavelength reference for precise RV measurements.

\term{Time of periastron passage} in the context of exoplanets refers to the specific moment $t_p$ when an exoplanet is closest to its host star during its orbit.

\term{True anomaly} is an angle between a planet's position at periastron (the point in its orbit closest to the star) and its current position, measured from the focus of the orbit.

\end{glossary}

\begin{glossary}[Nomenclature]
\begin{tabular}{@{}lp{34pc}@{}}
AGN & Active Galactic nucleus\\
au & Astronomical Unit \\
BD & Brown Dwarf \\
CARMENES & Calar Alto high-Resolution search for
M dwarfs with Exo-earths with Near-infrared and optical \'{E}chelle Spectrographs \\
ESPRESSO & \'{E}chelle SPectrograph for Rocky Exoplanets and Stable Spectroscopic Observations \\
ESO & European Southern Observatory \\
ELT & Extremely Large Telescope \\
F-P & Fabry-P\'{e}rot etalon\\
FOV & Field Of View \\
HARPS & High Accuracy Radial velocity Planet Searcher \\
HIRES & HIgh Resolution \'{E}chelle Spectrograph \\
HZ & Habitable zone \\
I$_2$ & Iodine molecule \\
JWST & James Webb Space Telescope \\
LFC & Laser Frequency Comb \\
MCMC & Markov chain Monte Carlo \\
MS & Main Sequence \\
MMR & Mean Motion Resonance \\
NIR & Near-Infrared \\
PSF & Point Spread Function \\
RV & Radial Velocity \\
SNR & Signal-to-Noise Ratio \\
Th-Ar & Thorium Argon \\
UVES & Ultraviolet and Visual \'{E}chelle Spectrograph \\
VLT & Very Large Telescope \\

\end{tabular}
\end{glossary}


\section{Introduction}\label{chap1:sec1}

The Doppler method has proven to be an indispensable technique for astronomers to precisely measure the kinematics of celestial objects such as stars, galaxies, and even the expanding Universe. The Doppler principle postulates that the frequency of waves changes depending on
the relative motion between the wave source and the observer measured along the light of sight.
In Astronomy, the Doppler method measures the wavelength shift of distinct spectral
features in emission, absorption, or both caused by well-understood physical phenomena.
In the context of stellar spectroscopy, we measure the dispersed electromagnetic intensity, where the 
Doppler effect is described in terms of the spectral absorption line shift in wavelength due to the motion of stars with respect to the observer.
Neglecting the relativistic effects\footnote{Relativistic effects are important when studying objects moving at very high velocities near the speed of light, such as observation of
high-redshift quasars, AGNs, and supernovae, which are key for testing the Hubble constant and
the rate of the expanding Universe.}, the basic formula used to calculate the stellar
spectral-line shift in wavelengths is:

\begin{align}\label{chap1:eq1}
\lambda' = \lambda \left(1 + \frac{v_r}{c}\right)
\end{align}

\noindent
where, $\lambda'$ is the measured wavelength, $\lambda$ is the wavelength emits in the rest
frame, $v_r$ is the velocity of the star relative to the observer, and $c$ is the speed of light.
The shift of the stellar electromagnetic radiation spectra could be recorded and analyzed,
and the line of sight (LOS) radial velocity component with respect to the observer is measured by simply:

\begin{align}\label{chap1:eq1}
v_r = c \left(\frac{\lambda' - \lambda}{\lambda}\right)
\end{align}

\begin{figure}[t]
\centering
\includegraphics[width=.995\textwidth]{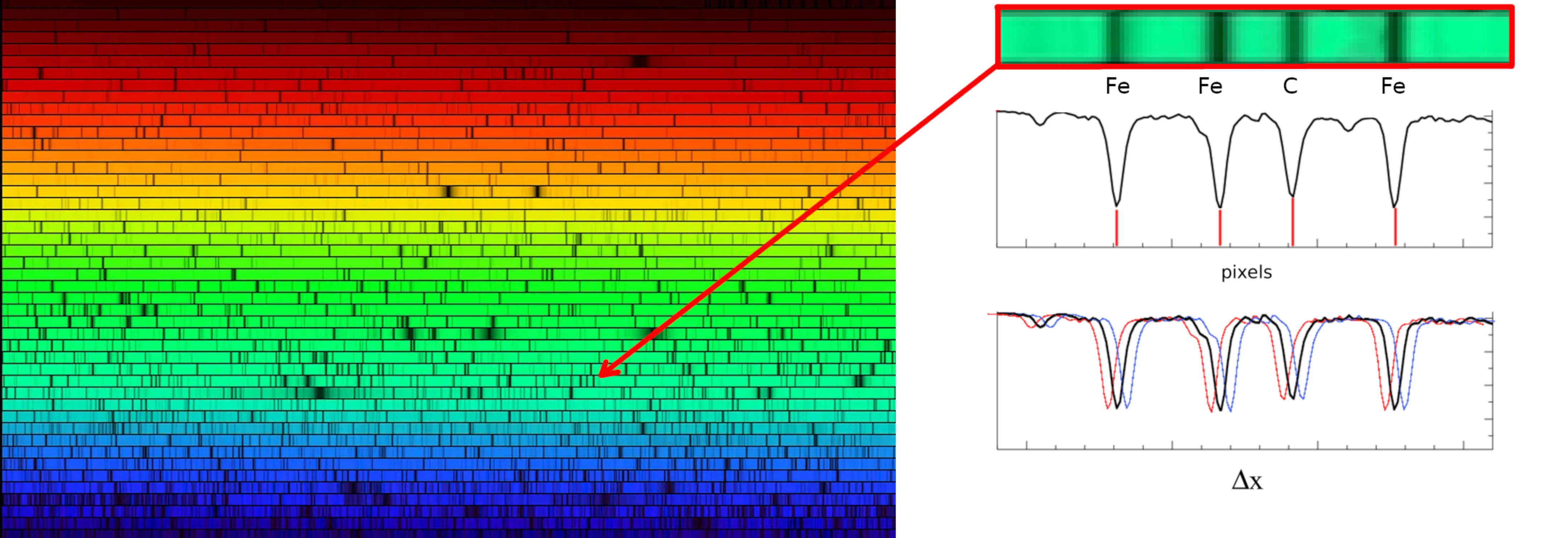}
\caption{Solar spectrum ranging from 392 nm (blue) to 692 nm (red), observed by the Fourier Transform Spectrograph at Kitt Peak National Observatory, Arizona, USA. The zoomed panel shows an example of
absorption lines recorded on the CCD detector with well-known wavelengths.
Measurements of red- and blue-Doppler shifts, crucial for detecting exoplanets, are achievable at the sub-pixel level.
}
\label{fig1}
\end{figure}

\noindent
Figure \ref{fig1} shows the structure of the optical spectrum of the Sun, which is a main
sequence (MS) star of class G2\,V. The Solar spectrum is not continuous across wavelengths
and reveals distinct absorption lines caused by the presence of various chemical elements
in the Solar atmosphere, such as light, and heavier iron-group elements like Carbon (C), Calcium
(Ca), Chromium (Cr), Manganese (Mn), Magnesium (Mg), Cobalt (Co), Nickel (Ni), and Iron
(Fe), among others. We know the exact wavelength positions of these spectral absorbers with
extreme accuracy in the laboratory rest frame. Similarly, for other stars, we record their
spectra and precisely measure the relative red- or blue-wavelength shift to determine the radial velocity (RV) of the object relative to the observatory.
Thus, if a sub-stellar companion, such as an extrasolar planet (exoplanet), or Brown Dwarfs (BD) orbits a star, it induces the star
to undergo a reflex motion around the system's center of mass (barycentre). The RV signal induced by an orbiting companion is very characteristic of and can be fully described by a
Keplerian model. The RV signal amplitude and shape depend on the companion's mass and
orbital geometry. For exoplanets, these Doppler shifts are rather small yet detectable by
using modern spectrographs and have led to the indirect detection of over 1,100 exoplanets to date.

\begin{BoxTypeA}[chap1:box1]{Important note}
\section*{On ``absolute'' and ``relative'' Radial Velocity measurements}
Each star has a unique direction of movement  (e.g., in the Galaxy) relative to the observer analyzing stellar spectra. Consequently, each star will exhibit a Doppler shift, resulting in a consistent "absolute" radial velocity. In the context of detecting exoplanets, we are interested in measuring the differential, or hereafter, the "relative" radial velocities of stars with respect to their barycentre, rather than their absolute velocities. Absolute RVs are not important for orbit determination and are naturally subtracted during the Keplerian modeling.
\end{BoxTypeA}

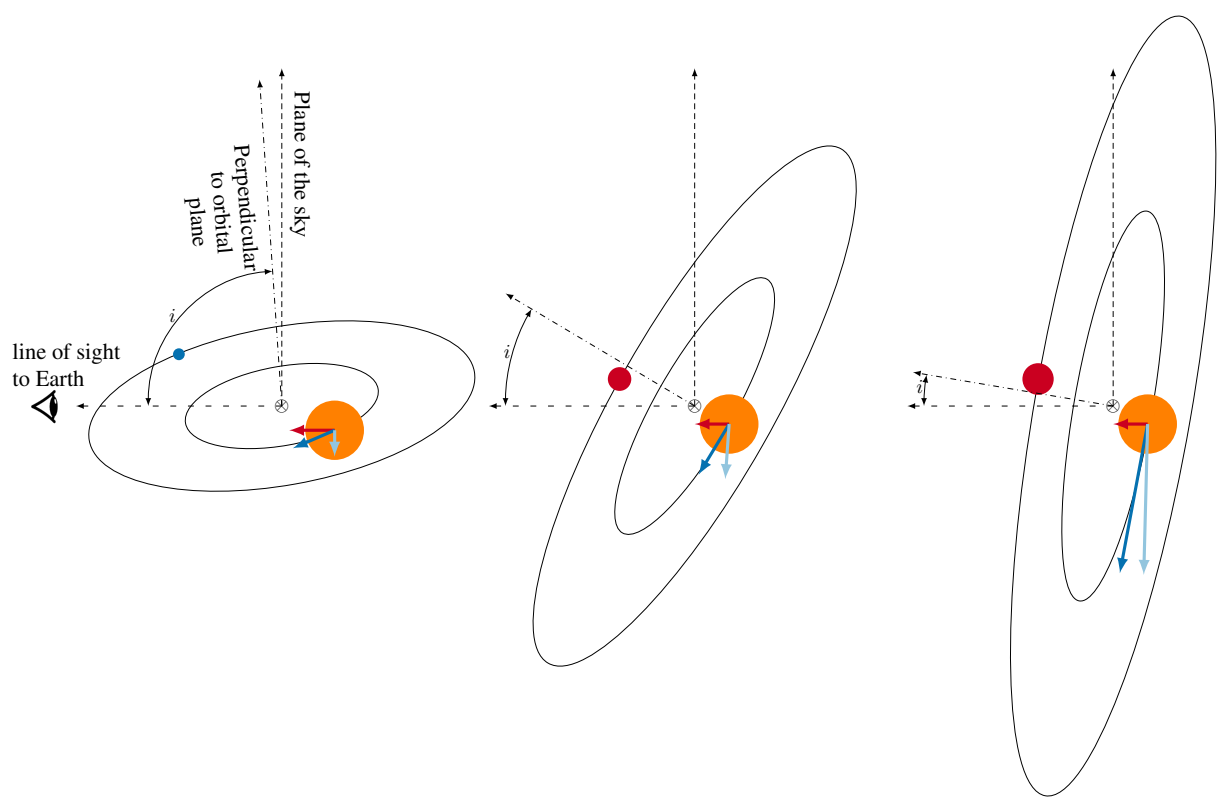
\begin{figure}[t]
\centering
\textbf{\large Three different stellar-substellar companion systems, each with a different line-of-sight inclination}\\

\includegraphics[width=.33\textwidth]{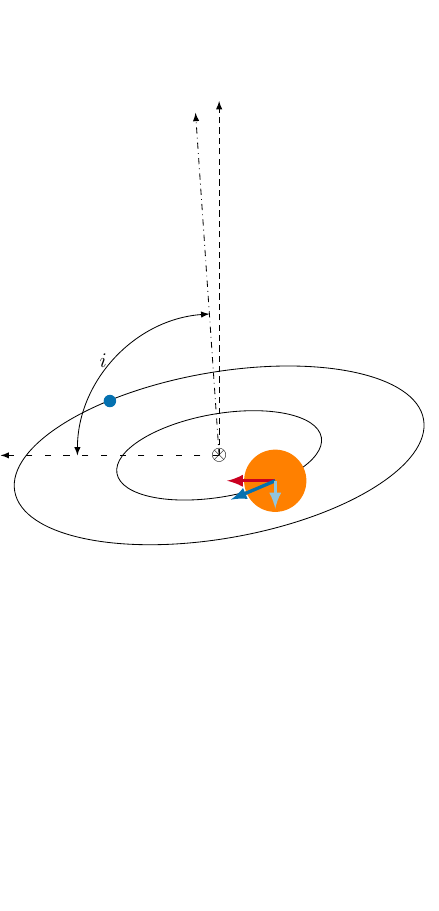}\put(-175,155){\includegraphics[width=0.03\textwidth]{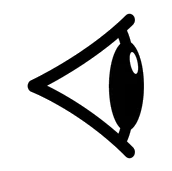}} \put(-180,180){\normalsize line of sight}\put(-180,170){\normalsize to Earth } \put(-75,280){\rotatebox{-90}{Plane of the sky}} \put(-98,260){\rotatebox{-85}{Perpendicular}}\put(-105,250){\rotatebox{-85}{to orbital}}\put(-115,240){\rotatebox{-85}{plane}}
\includegraphics[width=.33\textwidth]{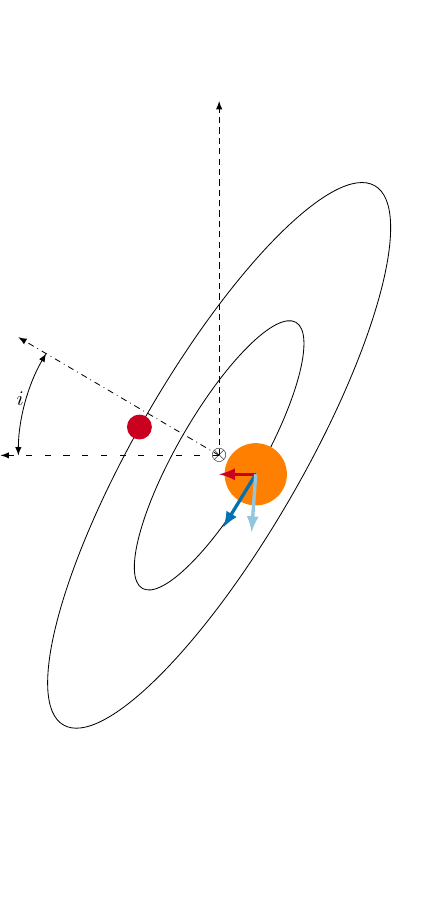}
\includegraphics[width=.33\textwidth]{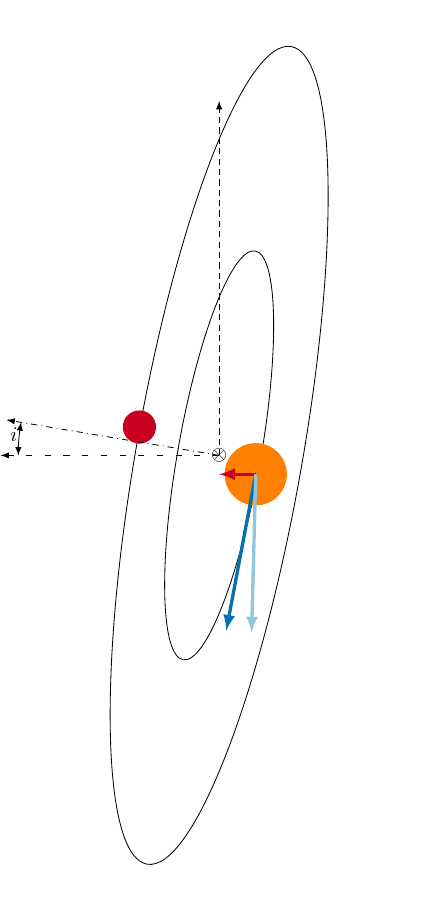}

\caption{The Doppler method measures the gravitational influence of an orbiting body (blue and red circle) on its host star (orange circle), detectable as a shift in the star's spectral lines due to the radial component of the star's motion towards or away from the observer.
While the star follows a Keplerian orbit influenced by its companion, we can only measure the component of this motion that is along the line of sight--the red vector, representing the radial velocity. The tangential velocity (light blue vector) and the true space velocity (blue vector) components of the star's velocity does not affect the spectral line shifts and thus remains undetectable using the
Doppler method. Since the inclination of the system is generally unknown without additional
observations, the Doppler method provides only a minimum mass estimate of the orbiting body, 
expressed as $m_p \sin(i)$, where $m_p$ is the mass of the companion, and $i$ is the 
inclination of the orbit to the line of sight.
 }
\label{fig2}
\end{figure}

Figure \ref{fig2} shows a simple schematic of the Doppler method. To discover stellar or sub-stellar companions, such as a BD or an exoplanet, all we need 
to do is regularly obtain stellar spectra and measure their Doppler shift to determine the stellar
radial velocities. From the characteristics of the RV signal as a function of time, a Keplerian 
model can reveal the spectroscopic parameters of the system, such as the orbital period and the 
orbital geometry of the perturbing object.
Figure \ref{fig2} also highlights the main disadvantage of the RV method. Wavelength shifts 
only allow us to measure the magnitude of the radial component of the stellar velocity (the red 
vector), which is along the LOS, while the tangential component of the velocity (the 
blue vector) remains unknown.
With the Doppler method, we cannot determine the orbital inclination of the system, and therefore, we can only estimate the minimum mass of the perturbing companion (i.e., the exoplanet), not their true dynamical 
mass. Figure \ref{fig2} illustrates that more massive but more inclined objects could, in 
principle, produce the same RV component vector. While the method itself seems fairly simple, as we will demonstrate further in this work, the precision RV measurements necessary to detect 
exoplanets are very complex and challenging.

\subsection{History of the Stellar Doppler Technique in Astronomy}\label{chap1:subsec1}

The first measurements of stellar radial velocity were done towards the end of the 19th century by the German astronomer Herman Carl Vogel, who used photography to record stellar spectra. Vogel's approach of recording spectra on photographic plates allowed researchers
to systematically measure stellar RVs with a typical precision of a few km\,s$^{-1}$, which
was sufficient to identify binary stars and stellar rotations. Systematic RV work has
been performed at the Lick Observatory, which provides an extensive collection of Doppler
measurements for bright stars, primarily observed between 1896 and 1926 using the Mills
Spectrographs attached to the 36-inch refractor at Mount Hamilton \citep{Campbell1926}.
Figure \ref{fig3} shows such an example of archival RVs taken with photo plates from Lick
and Victoria for the SB1 system HR\,6388. The RVs could be measured only for the primary
star, which is a K3\,III giant star, thus far more luminous than the secondary companion.
The photographic plate RVs have standard errors close to 1 km\,s$^{-1}$.
In the 1970s \citet{Griffin1978} complemented the HR\,6388 observations by taking RVs
using a photoelectric detector, slightly improving the RV precision, and leading to the firm binary orbit determination.

By the mid-20th century, astronomers managed to systematically record RVs of over 15,000 bright stars using photographic plates. Many of these targets were revealed as single (SB1) or double-line (SB2) binaries, and their orbital configurations were calculated based on the obtained 
RVs. Around that time, the visionary astronomer Otto Struve realized that precision was the 
only limit to detecting exoplanets. In the same way, binaries were uncovered, Struve postulated 
that RV measurements of a few hundred m\,s$^{-1}$ would be sufficient to detect close-in planetary companions with semi-major axes of about 0.02\,au, and masses similar to that of 
Jupiter. Otto Struve predicted a class of exoplanets now known as Hot-Jupiters (HJs) decades before the actual discovery of HJ exoplanets like 51 Peg\,b \citep{Mayor1995}. Furthermore, 
Struve's proposal for precision RV work for detecting exoplanets formulated the possibility of 
detecting exoplanet ''eclipses,'' now known as the transit detection technique for exoplanets. 
His visionary ideas marked significant foresight into methods that would only be realized much later.

Early efforts in stellar RV measurements faced challenges, particularly regarding precision. Techniques evolved over time, with notable contributions from researchers like \citet{Griffin1967}, who proposed strategies for improving Doppler precision by using 
photomultiplier detectors to scan the spectra along the direction of dispersion and calculate photoelectric RV measurements, reaching a steady precision of approximately 1\,km\,s$^{-1}$. 
Subsequent advancements, such as the development of gas cells \citep{Campbell1979} and 
simultaneous calibration with Thorium-Argon (Th-Ar) lamps \citep{Baranne1996}, further refined RV measurement techniques, leading to unprecedented levels of precision by the early 1990s.

\subsection{The Exoplanet revolution}\label{chap1:subsubsec1}

Early Doppler surveys of exoplanets began with the observations by \citet{Campbell1988},
who systematically monitored 16 stars for exoplanetary companions. They employed a hydrogen
fluoride (HF) gas cell as a wavelength calibration source \citep{Campbell1979}, achieving
an unprecedented precision at the time of approximately $\sim$ 13 m\,s$^{-1}$. This
precision was sufficient to detect Jovian-mass planets, but the limited sample size
prevented definitive detections. Nonetheless, \citet{Campbell1988} noted that seven stars
exhibited evidence of long-term, low-level variations. Among these, the K3 III giant star $
\gamma$ Cep\,A displayed RV variations, which could be potentially induced by a planetary
companion. Subsequently, \citet{Walker1992} dismissed this interpretation, attributing
the RV signal to the stellar activity of the red giant, influenced by theoretical prejudices
against the existence of short-period giant planets. The existence of the exoplanet $
\gamma$ Cep A b was only confirmed nearly two decades later by \citet{Hatzes2003}, who
verified that the RV variations of the primary binary component were indeed caused by a 1.7
M$_{\rm Jup}$ exoplanet with an orbital period of $\sim$2.5 yr.

In the late 1980s, \citep{Latham1989} identified ``A probable brown dwarf'' orbiting the
star HD\,114762. With a period of approximately 84 days and a minimum mass of about
M$_{\text{Jup}}$, this object exhibits characteristics of what we now call warm massive Jovian planets. Throughout the early 1990s, astronomers such as \citet{Hatzes1993} observed
long-period RV variations in giant stars, suggesting the presence of
giant planetary companions. These findings were later confirmed, notably for the star $
\beta$\,Gem, solidifying the role of RV measurements in detecting exoplanets.

The discovery of a 'hot Jupiter' exoplanet around the solar-type star 51\,Peg by \citet{Mayor1995} marked a revolutionary turning point\footnote{In 2019, 
Michel Mayor and Didier Queloz were awarded the Nobel Prize in Physics for this epochal discovery.} in the exoplanet search.
Utilizing the ELODIE \'{E}chelle spectrograph \citep{Baranne1996}, mounted on the 1.93\,m telescope at Observatoire de Haute Provence, pioneers Michel Mayor and Didier Queloz
captured an unambiguous RV signal. This signal was consistent with a
close-in giant planet orbiting its host star every 4.2 days. The minimum mass of the exoplanet was estimated to be 0.42 M$_{\rm Jup}$, challenging existing models based on our
Solar System and confirming the existence of the types of planets proposed by Struve four
decades earlier. The left panel of Figure \ref{fig4} displays the RV signal of 51\,Peg,
constructed from the original ELODIE RVs and supplemented by validation RVs obtained with
the Hamilton spectrograph \citep{Vogt1987} at Lick, USA by \citet{Marcy1997}, along with high-cadence observations from the HARPS spectrograph \citep{Mayor2003}. The discovery of
51 Peg\,b significantly expanded the search parameters for exoplanets, shifting the focus
to much shorter orbital distances and benefiting from improved precision in RV measurements.

Post-1995, the field of exoplanet research surged, with numerous astronomical groups
adopting the Doppler method to hunt for exoplanets, which led to a rapid increase in
exoplanet discoveries. Over the years, the precision of Doppler measurements has
dramatically improved, with many RV instruments now capable of detecting velocity changes
down to or even below 1 m\,s$^{-1}$. We can measure the velocity of stars many light-years
away with precision comparable to a normal human walking speed! This high precision allowed
astronomers to discover a planet in the habitable zone around our closest neighbor, Proxima
Centauri \citep{AngladaEscude2016}.

\begin{figure}[t]
\centering
\includegraphics[width=.48\textwidth]{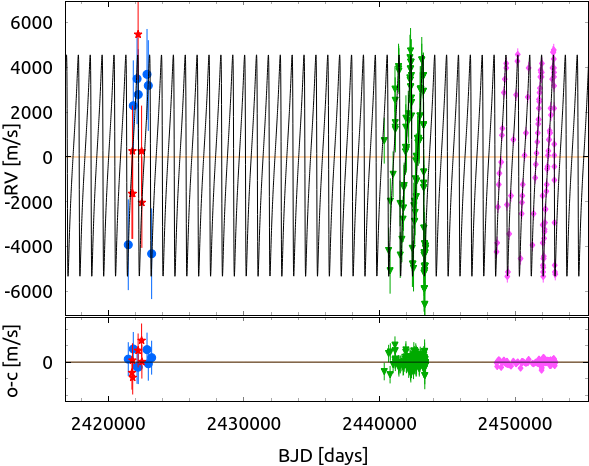}\put(-190,180){\normalsize 1918}
\put(-80,180){\normalsize 1975} \put(-25,180){\normalsize 2000} \hspace{0.5cm}
\includegraphics[width=.48\textwidth]{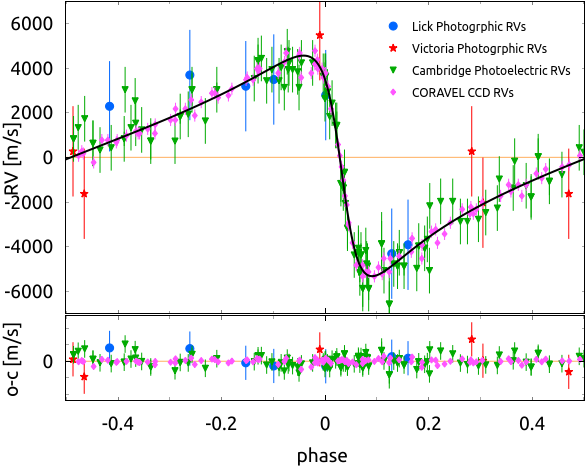}\put(-190,90){\large HR\,6388\,B} \put(-190,75){\large P = 876.0\,d}
\caption{RV measurements of the primary component of the SB1 binary HR\,6388, of which first RV measurements were taken over 100 years ago (the top axis lists years). The left panel shows the RV time series, illustrating the long-term observational campaign. The earlier RVs were obtained from Lick Observatory using photographic plates (blue), from Victoria using photographic RVs (red stars), more recent measurements by Cambridge using photomultiplier technology in the 1970s, and finally newer CCD-based RV observations taken with the CORAVEL spectrograph. The right panel displays the phase-folded model and RVs, revealing the large binary signal. The small panels below the main panels show the residuals (observed minus calculated, or o$-$c) of the model, highlighting the differences between the observed RV data and the model predictions. }
\label{fig3}
\end{figure}
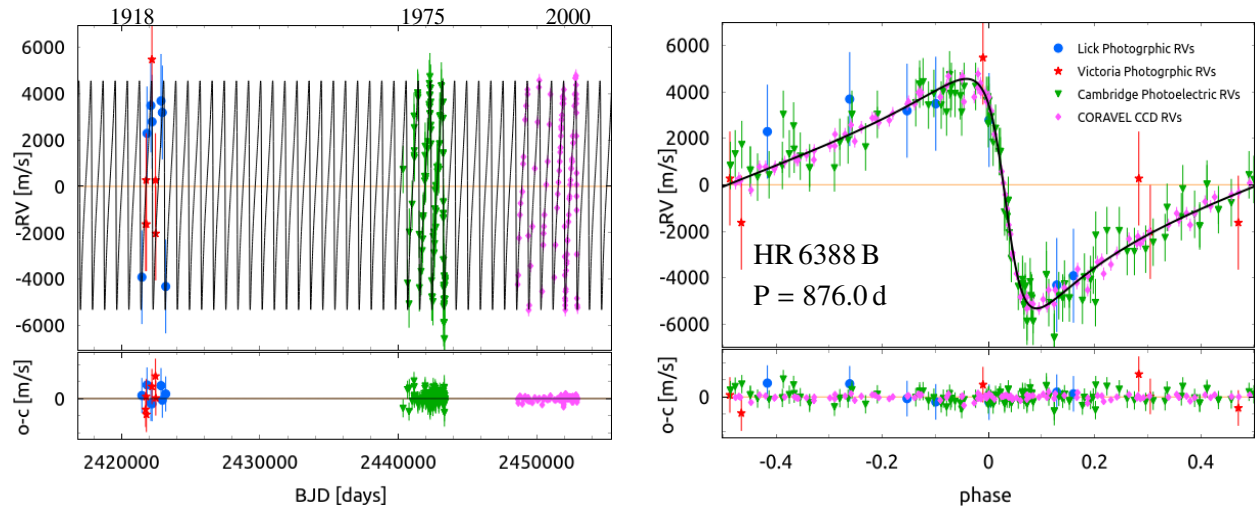

\begin{figure}[t]
\centering
\includegraphics[width=.48\textwidth]{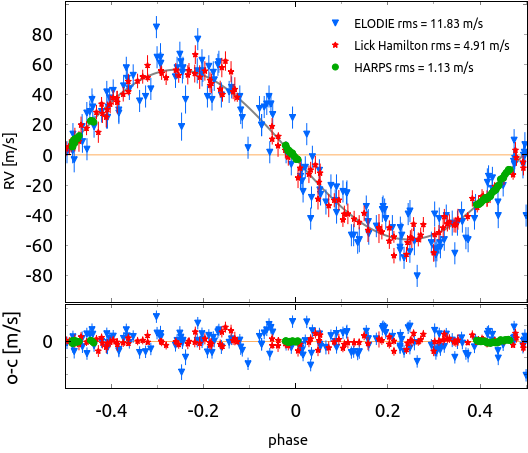}\put(-180,85){\Large 51 Pegasi\,b}  \put(-180,70){\large P = 4.2\,d}\hspace{0.5cm}
\includegraphics[width=.48\textwidth]{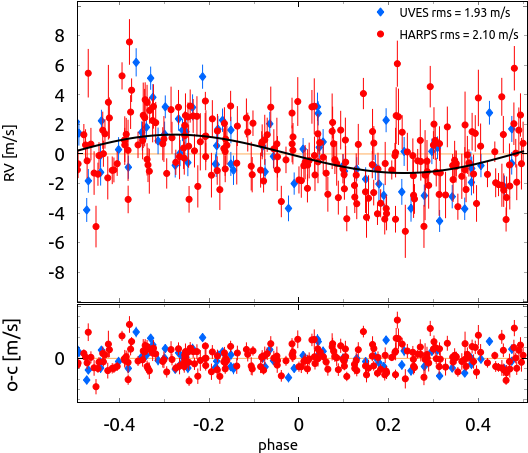}\put(-180,85){\large Proxima Cen\,b} \put(-180,70){\large P = 11.2\,d}
\caption{The left panel shows the phase-folded RVs of 51\,Peg, while the right panel shows the same for Proxima Cen, revealing the planetary-induced RV signal for these two stars. The top subplots list relevant information about the orbital period and the RV data used, along with their overall r.m.s. around the best fit. Same as in Fig.\ref{fig3}, the bottom subplots display the RV model residuals.}
\label{fig4}
\end{figure}

The right panel of Figure \ref{fig4} displays the RV signal of Proxima Cen\,b, detected
using the UVES and HARPS spectrographs. Note the dramatic evolution in precision from the
binary system HR\,6388 to the first planet discovered around a Solar-type star, 51 Peg\,b,
to our closest exoplanet neighbor, Proxima Cen\,b, as presented in Figures \ref{fig3} and
\ref{fig4}, respectively. The Stellar parameters and the best-fit models based on the RV
data for these three systems are summarized in Table \ref{tab:results}. Details on the
model parameters and methodology can be found in Sect. \ref{chap3}. The significant
increase in precision since the discovery of 51 Peg\,b has broadened the parameter space of
detectable exoplanets, enhancing our understanding of their masses and orbits.

\subsection{Impact of Doppler surveys for exoplanets}\label{chap1:subsubsec1}

After the discovery of 51 Peg b, the next twenty exoplanets were discovered using the precise RV method. These discoveries were made using targeted RV surveys for exoplanets, which brought about many more planets in the coming years.
In 1999, the first
transiting exoplanet discovered around the star HD\,209458 by \citet{Charbonneau2000},
marking a significant milestone in exoplanet research, providing a new method to study the
exoplanet demographics. To date, the search for extrasolar planets has yielded over 5500
confirmed planets, with more than 900 systems with multiple planets\footnote{up to-date
list is available on \url{https://expolanet.eu}}. Notably, the Doppler method of measuring
stellar RVs has led to the discovery of over 1100 exoplanets, while the transit photometry
technique, predominantly driven by the highly successful \textit{Kepler} space telescope
\citep{Borucki2010}, and recently the \textit{Transiting Exoplanet Survey Satellite}
\citep[\textit{TESS};][]{Ricker2015} has unveiled at least 4500 confirmed planets (which are usually validated with additional RV measurements!). The
remainder has been detected through direct imaging \citep[e.g.,][]{Marois2008},
gravitational microlensing \citep[e.g.,][]{Gaudi2012}, or other methods.

\begin{BoxTypeA}[chap1:box1]{Important note}
\section*{On transiting versus RV exoplanets}
A transiting exoplanet is detected when it passes in front of its host
star, causing a characteristic periodic dimming in the star's brightness,
which reveals the exoplanet's radius. Transit detection is only possible
for exoplanet orbits nearly aligned ($i \sim 90^{\circ}$) with the
observer's line of sight. Many transiting exoplanets have been discovered because the
method allows for the simultaneous observation of many targets with high photometric precision at a relatively low cost. In contrast, the RV method can detect Doppler shifts from both
transiting and non-transiting planets but is operationally expensive and limited to a few hundred bright stars per survey. The transit method reveals the orbital inclination $i$ but
is not sensitive to the companion's mass, which can lead to false-positive detections of
stellar companions. Therefore, knowing $i$, RV measurements are essential to validate the
planetary nature of transit events by measuring the companion's mass.
\end{BoxTypeA}

\begin{figure}[ht]
     \centering
     \includegraphics[width=0.48\textwidth]{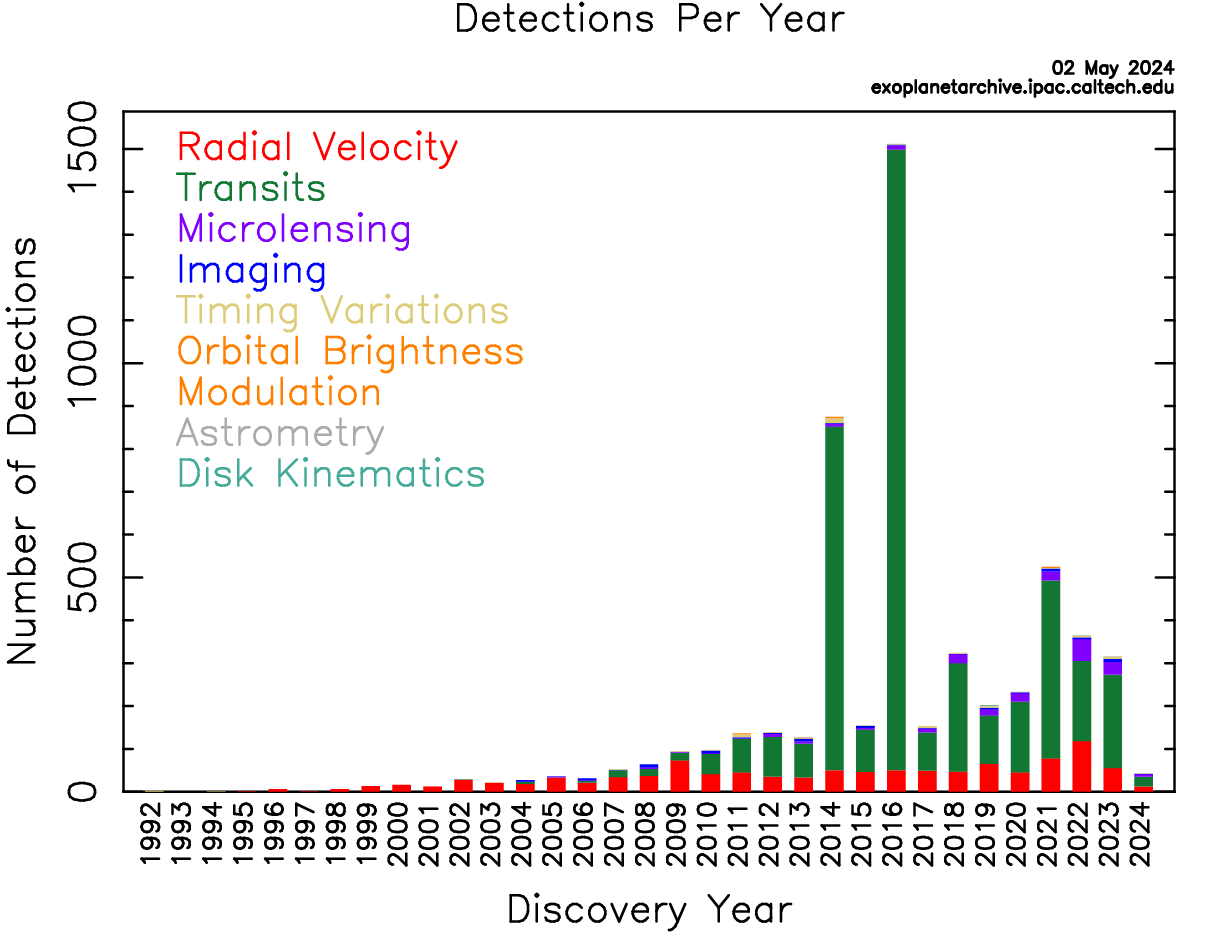}
     \includegraphics[width=0.48\textwidth]{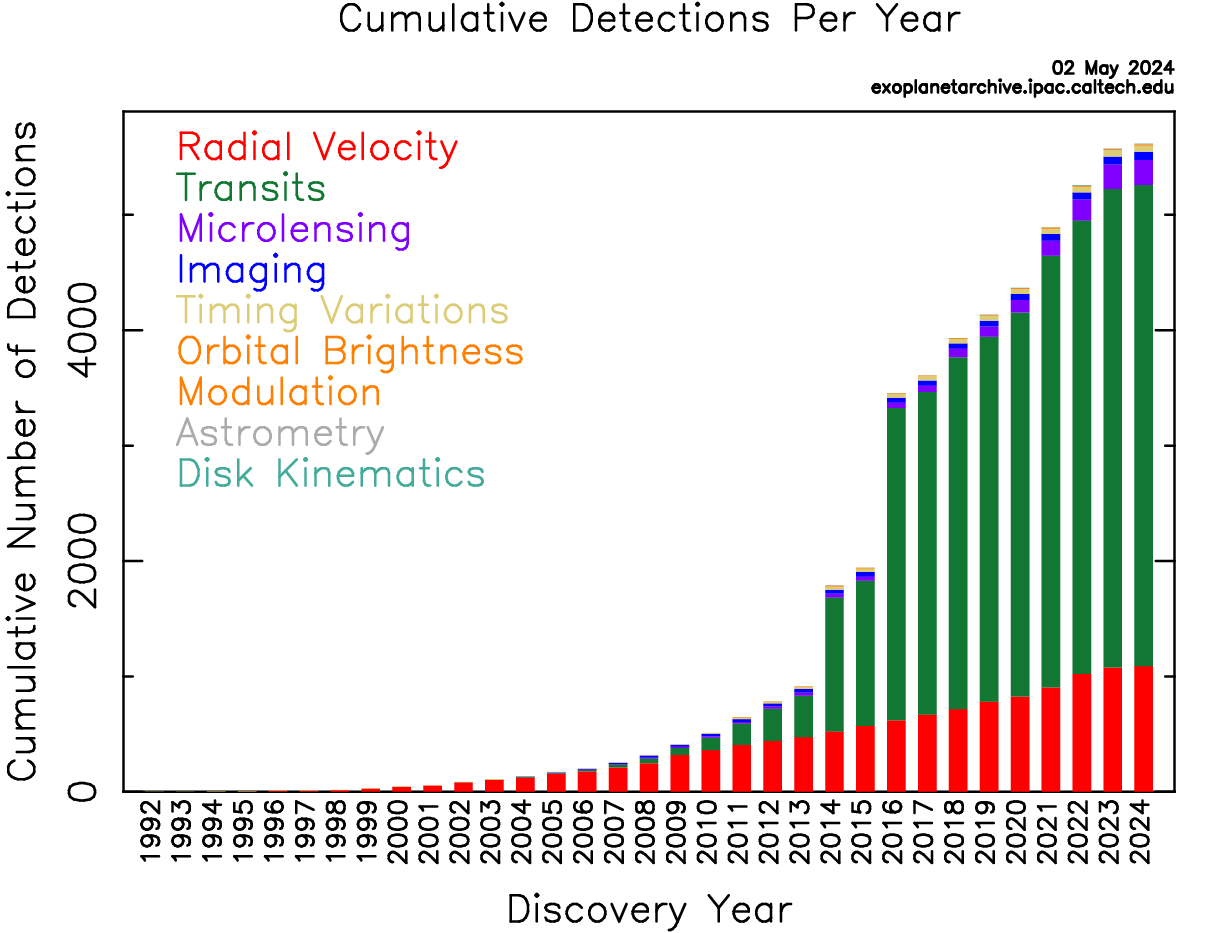}
     \caption{History of exoplanet detection. The left plot shows the number of exoplanet discoveries per year. The right plot shows cumulative discoveries in each year, divided into the different detection methods. Image credit: NASA Exoplanet Archive; Courtesy NASA/JPL-Caltech.}
     \label{fig5}
 \end{figure}

Figure \ref{fig5} presents a histogram of the exoplanet detection history by year (left) and cumulative (right), highlighting the progression and impact of detection techniques
over time. Currently, astronomers have several techniques to detect and characterize
exoplanets, but the two most efficient are the Doppler and the transit photometry
techniques. The RV method has had a steady detection rate over the years due to long-lasting, precise RV surveys. After 2010, the ground-based transit
monitoring for exoplanets also revealed exoplanets at an increased rate, surpassing the RV
detections. A few notable sharp peaks in the histogram are associated with the bulk of
exoplanets announced by the \textit{Kepler} and \textit{TESS} space telescopes.

\begin{figure}[t]
\centering
\includegraphics[width=.49\textwidth]{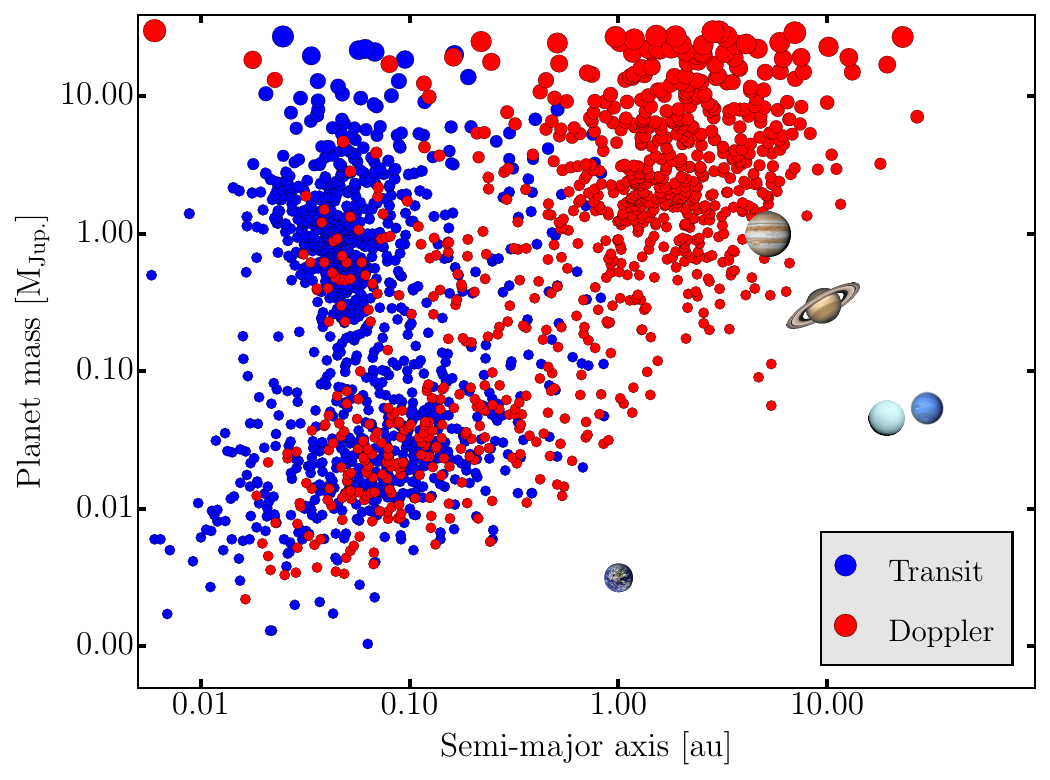}
\includegraphics[width=.49\textwidth]{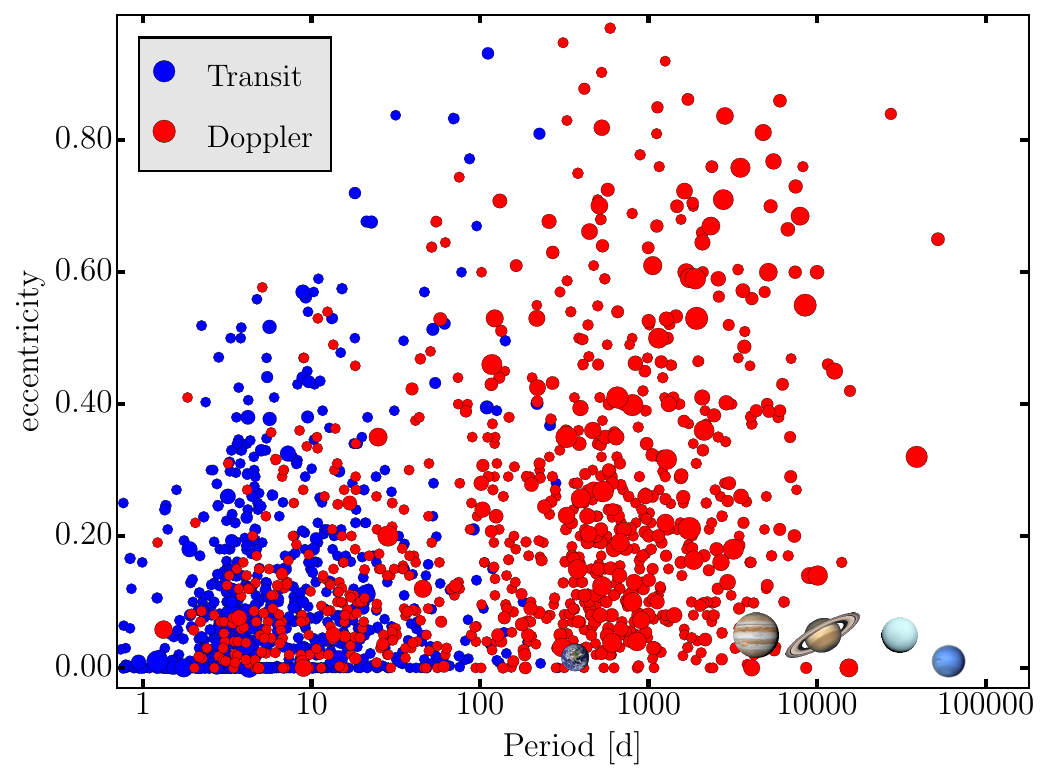}
\caption{Exoplanet demography revealed by the two most successful detection techniques: the Transit photometry (blue) and the Doppler RV (red) techniques. Larger symbols notate larger planetary mass. The left panel highlights detection biases favoring more massive and warmer planets, making solar system analogs and exo-Earths beyond reach at present days. The right panel, shows the period-eccentricity distribution, revealing a prevalence of exoplanets with moderate to high eccentricities, contrasting with the Solar system.
Data are collected from the ``Encyclopaedia of exoplanetary systems`` (\url{https://exoplanet.eu}).}
\label{fig6}
\end{figure}

The left panel of Figure \ref{fig6} illustrates the exoplanet demography by showing the distribution of exoplanet mass versus orbital distance as revealed by the Doppler and transit detection techniques. The distribution of exoplanets is significantly influenced by observational biases inherent to these methods.
Both techniques are exceptionally prompt at detecting more massive and shorter-period planets due to their higher detectability -- the larger and more massive the exoplanet, the greater the induced RV and transit signals. Improved precision over the past two decades has also enabled the detection of a plethora of short-period, low-mass exoplanets, ranging from rocky bodies slightly larger than Earth to mini-Neptunes, thus introducing a new class of exoplanetary bodies, which to a large degree came as a surprise to astronomers.
The Doppler method is sensitive to long-period colder massive companions similar to Jupiter, thanks to the long observational baselines of RV surveys. We have discovered a large number of these exoplanets, typically
more massive, but continuing the legacy RV surveys, we are on a quest to recover the occurrence rate of Saturn analogs in the near future. Both techniques still need a longer observational baseline and better precision for exoplanet analogs of Uranus, Neptune, and the inner solar system planets.

One of the ultimate goals in the exoplanet field is to determine the occurrence rate of Solar System analogs, particularly Earth-like planets capable of sustaining life. Achieving this goal requires an observational baseline on the order of decades with extremely precise RV measurements on the order of a few cm\,s$^{-1}$, which unfortunately remains beyond our reach to date. Consequently, we must rely on existing observational data to refine and calibrate our theories of planet formation, striving for a deeper understanding of these mechanisms and indirectly uncovering the observational diversity.

RV surveys have greatly enhanced our knowledge of exoplanet diversity and their potential formation processes. For instance, despite the high detection rate of HJs, we know that these objects are relatively rare, with only about 1\% of Solar-type stars hosting HJs, and an even lower fraction among lower-mass stars \citep{Sabotta2021}. Many RV-detected planets are found in multi-planet systems, providing valuable insights into their dynamical interactions, which trace their formation history and evolutionary paths. The existence of HJs and multiple-planet systems close to or locked in mean-motion resonance (MMR) has opened a new field in astrophysics, aiming to explain their formation and advancing the theories of convergent giant-planet migration \citep[e.g.,][]{Lee2002}.

The right panel of Figure \ref{fig6} depicts the distribution of exoplanet eccentricity versus orbital period. Measuring the eccentricity of
hundreds of exoplanets with RVs has been a crucial factor in understanding the planet's formation and evolution pathways.  The RV method, alone or in combination with transit observations, has revealed that many exoplanets possess moderate to highly elliptical orbits, contrasting with the nearly circular orbits observed in our Solar System. This observation is attributed to planet-planet scattering events during exoplanet evolution, suggesting some violent history of exoplanets \citep{Ford2008}.
Another critical output of evidence is that we owe the dedicated Doppler surveys the detailed planet occurrence rates as a function of the stellar host mass and metallicity \citep[e.g.,][]{Fischer2005,Wolthoff2022}, which suggests a positive correlation between stellar metallicity and the occurrence rate of planets. By providing detailed insights into planets' mass, orbit, eccentricity, and period, Doppler surveys have enriched our comprehension of the dynamic architectures of planetary systems, highlighting their diversity and complexity.

\begin{table*}
  \centering
  \caption{Best fit orbital parameters, minim masses, and semi-major axis alongside their 1$\sigma$ uncertainties for the RV model fit presented in Figures \ref{fig3} \& \ref{fig4}.
  These are for systems studied in this work; HR\,6388, 51\,Peg\,b, and Proxima\,Cen\,b, using
  archival and literature RVs and modeled with the Exo-Striker tool.}
  \begin{tabular}{ p{4.3cm} p{4.3cm}  p{4.3cm} l}
  \hline  \noalign{\vskip 0.7mm}
       Primary           & HR\,6388 A  & 51 Peg & Proxima Cen$^{a}$\\
  \hline \noalign{\vskip 0.7mm}

 Sp. type  & K3\,III    & G2\,IV & M5.5\,V\\ \noalign{\vskip 0.9mm}
 Mass (M$_\odot$) & 1.02  & 1.11 & 0.12 \\ \noalign{\vskip 0.9mm}
 Radius (R$_\odot$) & 18.8  & 1.27 & 0.14\\ \noalign{\vskip 0.9mm}
 [Fe/H] (dex) & -0.17  & 0.2 & -0.04 \\ \noalign{\vskip 0.9mm}
 T$_{\rm eff.}$ (K) & 4335  & 5793 & 3306 \\ \noalign{\vskip 0.9mm}

 distance (pc) & 82.04  & 15.53 & 1.302 \\ \noalign{\vskip 0.9mm}

  \hline \noalign{\vskip 0.7mm}
       Companion parameters  & HR\,6388 B  & 51\,Peg\,b & Proxima Cen\,b \\

  \hline \noalign{\vskip 0.7mm}

$P$ (d) & $876.3_{-0.1}^{+0.1}$              & $4.2308_{-0.0001}^{+0.0001}$ & $11.184_{-0.001}^{+0.001}$ \\ \noalign{\vskip 0.9mm}
$K$ (m\,s$^{-1}$) & $4915.8_{-39.7}^{+40.0}$ & $56.4_{-0.4}^{+0.4}$ & $1.25_{-0.16}^{+0.16}$ \\ \noalign{\vskip 0.9mm}
$e$  & $0.621_{-0.005}^{+0.005}$ & $0.008_{-0.005}^{+0.006}$ & $0.035_{-0.02}^{0.03}$  \\ \noalign{\vskip 0.9mm}
$\omega$ (deg) & $97.2_{-0.9}^{+0.9}$ & $304.4_{-16.0}^{+29.4}$ & $259.0_{-89.7}^{+64.3}$ \\ \noalign{\vskip 0.9mm}
$Ma$ (deg) & $45.2_{-1.2}^{+1.3}$ & $324.7_{-29.5}^{+15.9}$ & $207.0_{-64.6}^{+89.7}$ \\ \noalign{\vskip 0.9mm}
Mass (m $\sin i$)  & $0.197_{-0.002}^{+0.002}$ (M$_{\odot}$) & $0.480_{-0.003}^{+0.003}$ (M$_{\mathrm{Jup}}$) & $1.06_{-0.13}^{+0.13}$ (M$_\oplus$)\\ \noalign{\vskip 0.9mm}
$a$ (au) & $1.914_{-0.001}^{+0.001}$ & $0.0530_{-0.0001}^{+0.0001}$ & $0.04828_{-0.0001}^{+0.0001}$ \\ \noalign{\vskip 0.9mm}

Epoch  & 2421463.710 & 2449610.527 & 2451634.731 \\ \noalign{\vskip 0.9mm}

  \hline \noalign{\vskip 0.5mm}
  \end{tabular}{%
   \begin{tablenotes}
\item[a]{$a$ - Note that at least one more confirmed exoplanet in the Proxima Cen system on shorter orbit \citep{Faria2022}, and an additional long-period candidate exoplanet \citep{Damasso2020} which signals were not modeled in this work.}
 \end{tablenotes}
 }

      \label{tab:results}
\end{table*}

\noindent
\vspace{10pt}
\rule{\textwidth}{2pt}
\vspace{5pt}

\section{Measuring precise Radial Velocities}
\label{chap2}

Doppler surveys utilize high-resolution spectroscopy to measure the small periodic wobbles in the motion of stars, which result from the gravitational pull of orbiting planets. The development of sensitive spectrographs, coupled with the stable wavelength calibration of spectra recorded on Charge-Coupled Devices (CCDs), and precise RV extraction numerical routines are critical for the precision of RV measurements. In this section, we introduce the fundamental principles required to measure precise RVs.

\subsection{Modern \'{E}chelle spectrographs}
\label{chap2.1}

Detecting the bulk of exoplanets through precision RVs has only become feasible thanks to the significant technological advancements over the past half-century. A major turning point in precision RV work was the introduction of CCDs as alternatives to photo plates and photomultiplier detectors. The highly efficient CCDs enabled the digital recording of high signal-to-noise ratio (SNR) spectra, allowing for the calibration and analysis of data using numerical techniques. However, classical long-slit spectrographs were not ideally suited for precision RV exoplanet hunting due to their limited wavelength coverage and relatively modest spectral resolution.

\begin{figure}[ht]
    \centering

   \includegraphics[width=0.9\textwidth]{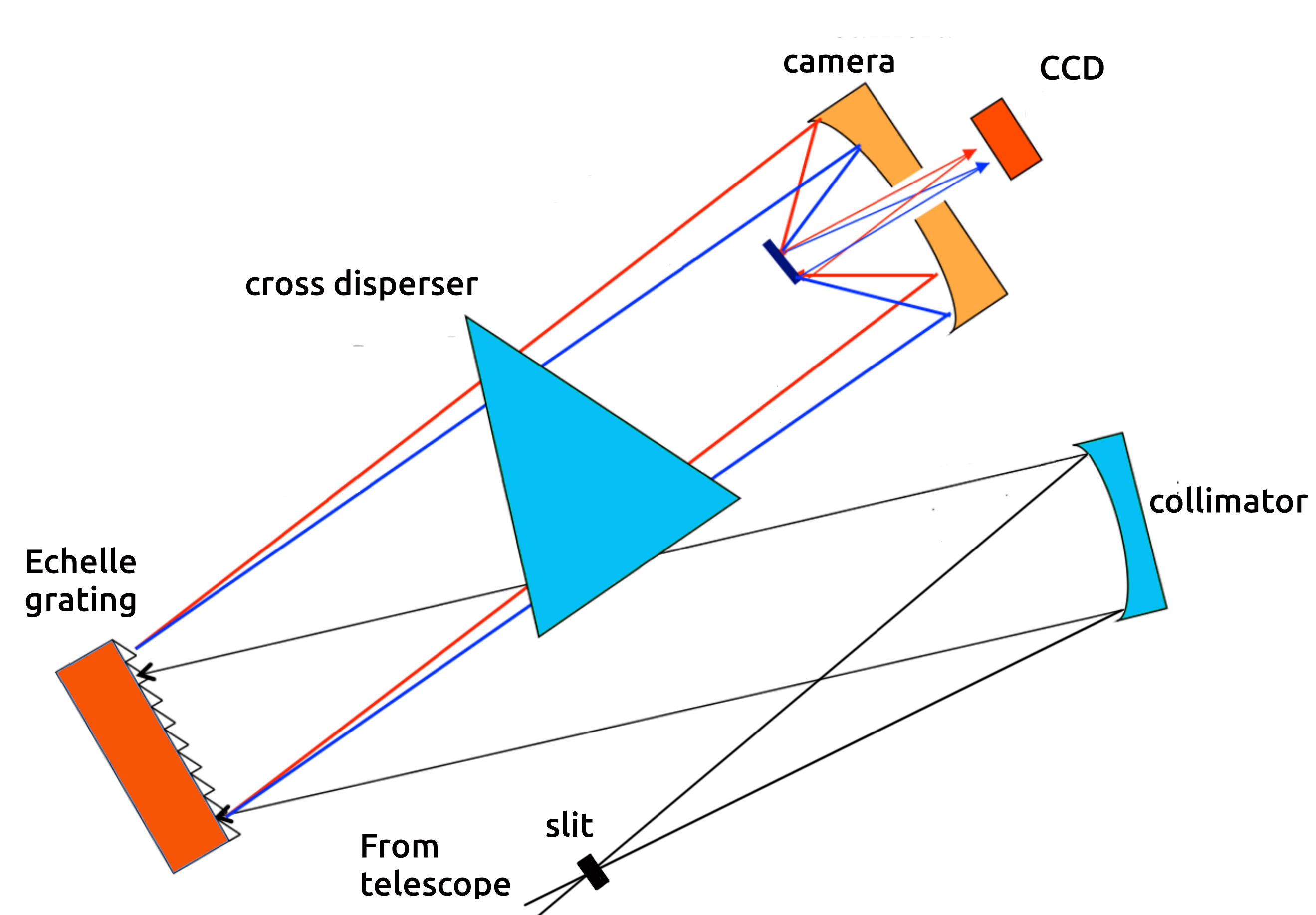}

\caption{A simplified diagram of an \'{E}chelle spectrograph. In addition to the grating dispersive element, there is a second cross-disperser optical element, which is usually a prism.}

    \label{fig:spectroES}
\end{figure}

To effectively use stellar photons, astronomers and optical engineers needed to expand the wavelength coverage and resolution to record more spectral lines simultaneously on the CCD. This led to the brilliant concept of \'{E}chelle\footnote{\'{E}chelle comes from the French word for "ladder."} spectrographs, which provide high resolving power $R$ over a very large wavelength range. The resolving power $R$ is defined as:

\begin{equation}
R = \frac{\lambda}{\Delta \lambda}
\end{equation}

\noindent
where $\lambda$ is the wavelength of light, and $\Delta \lambda$ is the smallest difference in wavelength that the instrument can resolve.
This defines the instrument's ability to distinguish between two closely spaced spectral lines. \'{E}chelle spectrographs typically have $R \equiv \lambda / \Delta \lambda \sim 40\,000 - 120\,000$ and provide wavelength coverage in the region 380-900\,nm (which varies widely across different designs), where CCDs are most efficient.

Figure \ref{fig:spectroES} shows a simplified schematic of a modern \'{E}chelle spectrograph. Note that many design variants exist depending on the telescope's optical systems and the goals of the scientific programs. It is important for the reader to recognize that the \'{E}chelle spectrographs are considerably more complex than what is illustrated in Fig. \ref{fig:spectroES}.
More details on \'{E}chelle spectrograph concepts can be found in the dedicated book of \citet{Eversberg2015}.

\begin{figure}[ht]
    \centering
    \begin{tikzpicture}
        \node[anchor=south west,inner sep=0] (image) at (0,0) {\includegraphics[width=0.95\textwidth]{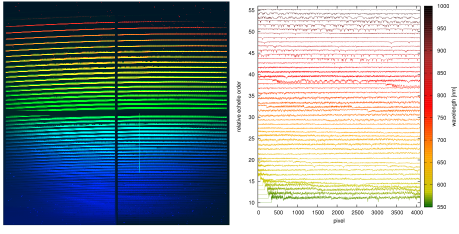}};
        \begin{scope}[x={(image.south east)},y={(image.north west)}]
             \coordinate (panel2ll) at (0.0,0); 
            \coordinate (panel2lr) at (0.5,0); 
            \coordinate (panel2ul) at (0.5,0.5); 
            \coordinate (panel2ur) at (0.0,1); 

            \draw[thick,->,red] (panel2ll) -- (panel2lr);
            \draw[thick,->,red] (panel2ll) -- (panel2ur);
            \node[below, red] at (0.25,0) {\'{E}chelle (main) dispersion};
            \node[rotate=90, above, red] at (0.0,0.5) {\large cross- dispersion};
        \end{scope}
    \end{tikzpicture}
\caption{The left panel  displays the resulted \'{e}chelle spectra on a CCD detector. The right panel shows the order-by-order wavelength calibrated 1D extracted spectra from which RVs are extracted. The spectra were taken for Luyten's star (GJ\,273) with the CARMENES spectrograph and provided by Adrian Kaminski and Mathias Zechmeister (priv. communication).}

    \label{fig:spectroES2}
\end{figure}
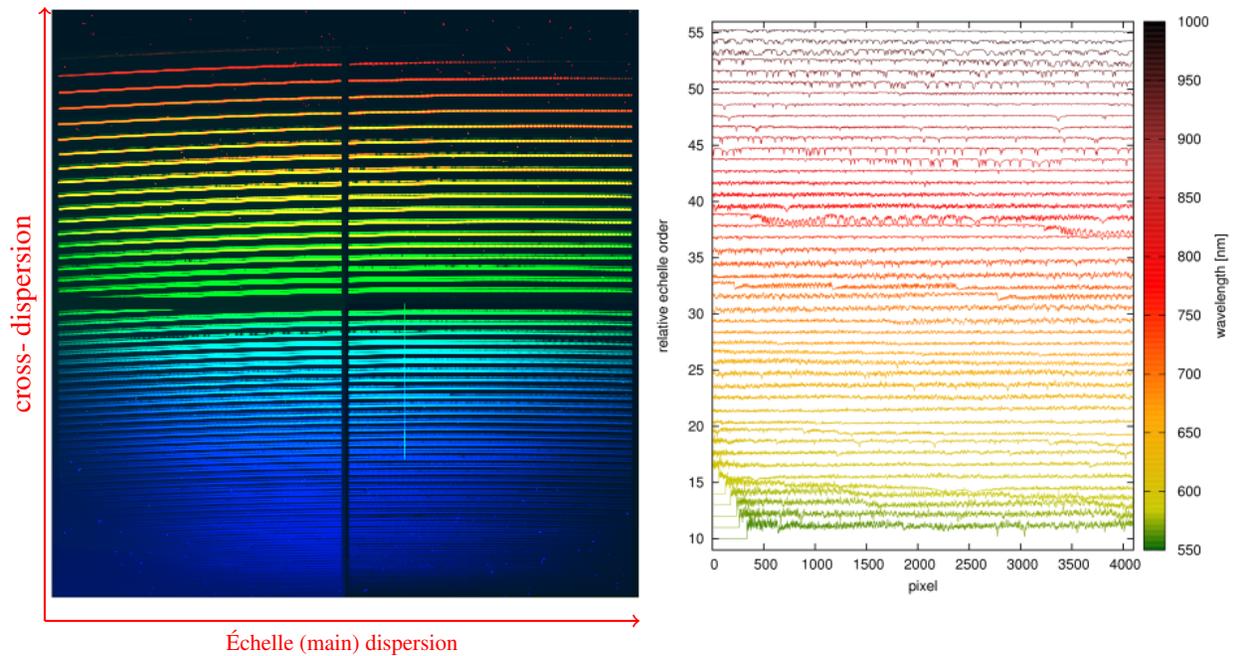

Briefly, an \'{E}chelle spectrograph has an entrance through which light from the telescope passes. Modern \'{E}chelle spectrographs are usually fed with an optical fiber, which sends the light from the telescope focus to the spectrograph. As we will discuss later, the spectrograph is usually placed in a separate temperature-stabilized room to ensure stability. Thus, the usage of a well-selected optical fiber is critical for the performance of the instrument. Telescope light from the fiber enters the spectrograph via an image slicer, which is an optical element that divides a light beam into narrower beams to increase the resolution and efficiency of a spectrograph, and thus is one of the components defining the resolution of the instrument. A collimator mirror with the same focal ratio as the telescope makes the light beam parallel to the first dispersing element; an \'{E}chelle diffraction grating. \'{E}chelle gratings have a relatively small grating constant and a large angle of incidence, resulting in many overlapping wavelength intervals in high orders \citep{Eversberg2015}. These overlapping orders are hardly useful. However, this ''defect`` could be converted to a useful ''effect``  by employing a second cross-disperser prism, which disperses the orders along the y-axis. Finally, the cross-dispersed orders are focused by a camera, which projects the spectra onto the CCD, filling most of the detector with useful spectra across many orders. This is
illustrated in the left panel of figure \ref{fig:spectroES2},
which shows a real spectrum (artificially colored, however)
from the Optical arm spectrograph of the CARMENES instrument \citep[see,][and reference therein]{Reiners2018b}. The right panel of figure \ref{fig:spectroES2}, shows the extracted orders into one-dimensional spectra, ready to be wavelength calibrated and the RVs extracted.

\noindent
To summarize, \'{E}chelle spectrographs offer:

\begin{itemize}
\item {\bf High spectral resolution} -- \'{E}chelle spectrographs can separate spectral features that are very close in wavelength, which is particularly important for precision RV work.

\item {\bf Large Wavelength Coverage} -- The broad wavelength range of \'{E}chelle spectrographs makes them highly efficient for capturing a wide array of astronomical phenomena in a single exposure. In the context of exoplanet studies, the extensive wavelength coverage provides a sufficient number of stellar lines to measure precise RVs. Additionally, it helps differentiate between signals caused by stellar activity and those caused by companions. Doppler shifts due to companions are wavelength-independent, whereas activity-induced shifts are not.

\item {\bf Compact design} -- generally, \'{E}chelle spectrographs are compact instruments for 3-m class telescopes, although this depends on many factors. Some \'{E}chelle spectrographs on 8-m class telescopes such as ESPRESSO take considerable area in the VLT's Combined Coud\'{e} Laboratory.
Only the vacuum tank alone is approximately 2 meters in diameter and about 4 meters in length, making it a large and sophisticated, {\it state-of-the-art} piece of equipment.

\item {\bf Versatility} -- given the above, these instruments can be used for a wide range of spectroscopic interests, depending on specific observational needs, as long as the task is feasible. For instance, HIRES at Keck Observatory in the USA is a general-purpose spectrograph that has led to the discovery of over a hundred exoplanets.
\end{itemize}

\noindent
Despite these advantages, \'{E}chelle spectrographs are complex and costly to construct and operate. Their data processing is also very challenging, requiring sophisticated techniques to separate and calibrate the overlapping spectral orders. Additionally, these complex instruments must maintain very stable temperature and mechanical conditions to achieve m\,s$^{-1}$ RV precision, as we will discuss in the following sections.

\subsection{Wavelength calibration}
\label{chap2.2}

Figure \ref{fig:spectroES2} illustrates that modern \'{E}chelle spectrographs record spectral intensity on a CCD detector. However, measuring Doppler shifts directly in the CCD pixel space is impossible. Each time we point the telescope at our stellar target to record its spectra, the absorption lines fall at different pixel positions. Moreover, stellar line shifts consistent with exoplanet signals occur at a sub-pixel level, but depending on the resolution of the spectrograph and the pixel size of the CCD, a typical shift of a single pixel can be transferred to a change in RV between 1 and 3 km\,s$^{-1}$. So, how are we expected to measure the Doppler shift down to m\,s$^{-1}$ precision? To achieve this, it is necessary to convert the relatively sparse pixel space to a finer grid of wavelengths. This is a standard technique achieved by calibrating our spectrograph with a stable calibration source. For exoplanet studies, the calibrator must remain consistent over many years to precisely measure the relative Doppler shift as a function of time. It is important to recall that achieving the desired RV precision ultimately relies on statistics by measuring the Doppler shift of as many stellar lines as possible in the calibrated wavelength space. Here, we discuss the two most common techniques for wavelength calibration, which are essential for detecting exoplanet signals.

\subsubsection{The I$_2$ cell method}
\label{chap2.2.1}

The I$_2$ cell method provides an efficient and inexpensive way to achieve
long-term precision RV measurements in general purpose \'{E}chelle spectrographs.
The application to stellar RV work was proposed by \citet{Marcy1992},
who demonstrated that placing an I$_2$ gas cell in the light path at the entrance of the spectrograph, provides many narrow and very well-defined I$_2$ spectral lines, which can be used as a wavelength reference.

\begin{figure}[t]
\centering
\includegraphics[width=.95\textwidth]{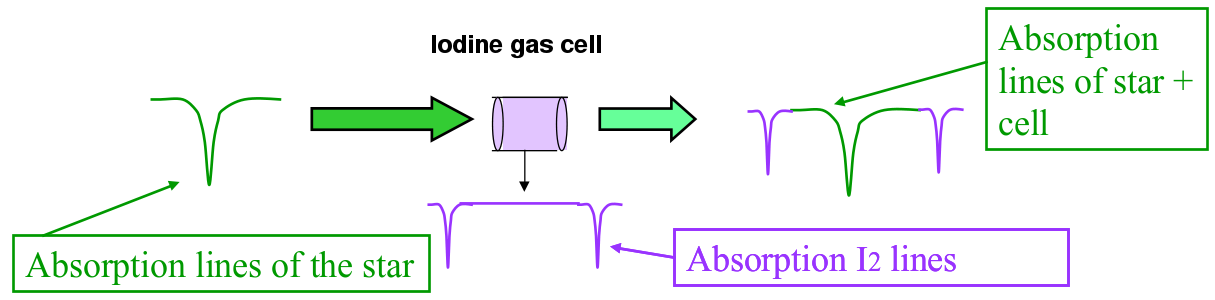}
\caption{The iodine cell method scheme. Light from the telescope passes through a glass chamber filled with I$_2$ gas. The absorption spectrum of the I$_2$ is superimposed onto the stellar spectrum, serving as a very stable wavelength reference. Credit: Figures are from Artie Hatzes's lecture notes (priv. communication), modified by the author.}
\label{fig32}
\end{figure}

Figure \ref{fig32} shows a simplified scheme of the I$_2$ cell method.
I$_2$ gas cells are favored for their rich spectrum of narrow absorption lines within the 500-600\,nm wavelength range, providing very stable reference lines superimposed on the stellar spectra. The I$_2$ cells offer practical benefits such as compact size, ease of use, and low maintenance. These devices do not require additional optics, making them an ideal upgrade for high-resolution general-purpose spectrographs whose primary goal was not initially exoplanet surveys.
This means that any suitable spectrograph could be transformed into a precision RV machine. The main requirement is to maintain a constant operating temperature near 50$^{\circ}$C.
The HIRES spectrograph was the first Doppler instrument capable of delivering precise RVs down to 3 m\,s$^{-1}$, or even better for dwarf stars \citep{Butler1996}.
Through ongoing instrument and RV extraction pipeline optimizations, precision has significantly improved over the years, reaching about 1\,m\,s$^{-1}$ for bright stars \citep{Butler2017}. Over the years, many other \'{E}chelle spectrographs were equipped with an I$_2$ gas cell for detecting exoplanets, proving to be very effective.

There are, of course, some operational challenges with the I$_2$ method. For instance, the data reduction and RV extraction processes are complicated. First, obtaining very high-resolution spectra of the I$_2$ gas cell and constructing a model template is required. Although this is needed only once, only a few facilities worldwide can scan spectra with a resolution (R $>$ 500,000), which is necessary for this task. Additionally, high SNR spectra without the I$_2$ spectra for each studied star must be obtained, known as a ''stellar template'', which consumes traditional operational time. The I$_2$ lines contaminate the stellar spectrum, complicating or even preventing further spectral analyses such as abundance studies or examining spectral line shapes needed for stellar activity analysis.

Despite these challenges, the I$_2$ method remains an excellent tool for studying the occurrence rate of Jovian analogs. This task does not require extreme cm\,s$^{-1}$ precision but does need a long-term baseline of stable RV reference, which is easily achieved with I$_2$ cells. Legacy and future RV surveys using spectrographs equipped with I$_2$ cells are relatively inexpensive to build and maintain, potentially revealing many more massive exoplanets in the coming decades.

\begin{figure}[t]
\centering
\includegraphics[width=.95\textwidth]{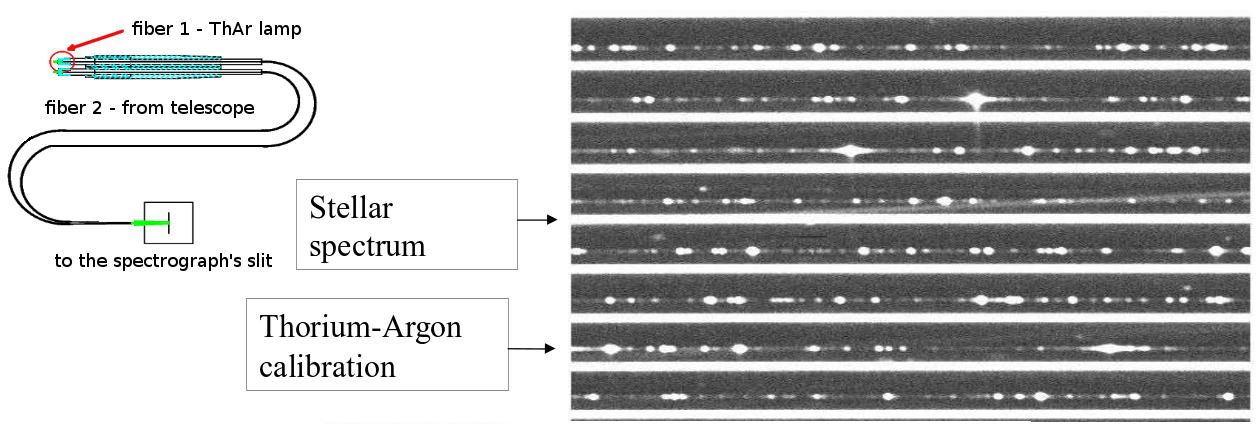}
\caption{A spectrum recorded with the HARPS spectrograph using a simultaneous Th-Ar calibrator lamp. Two optical fibers are illuminated simultaneously: one from the telescope and the other from the Th-Ar lamp. The seemingly continuous spectral orders are stellar spectra from the star fiber, whereas the Th-Ar spectrum consists of strong emission lines used for wavelength calibration. Credit: Figures are from Artie Hatzes's lecture notes (priv. communication), modified by the author.}
\label{fig:Th-Ar}
\end{figure}

\subsubsection{The simultaneous Th-Ar calibration method}
\label{chap2.2.2}

The Thorium-Argon (Th-Ar) lamps have been widely used as a wavelength calibration source in astronomy for many decades due to their advantages in spectral analysis. Thorium, serving as an excellent cathode, offers a dense array of narrow spectral lines across visible wavelengths, which helps in accurate wavelength calibration. A key technique involves recording the Th-Ar spectrum simultaneously with the stellar spectrum using optical fibers.  This setup involves using one fiber optic to channel light from the star and another for light from a calibration lamp to the spectrograph, where both spectra are recorded side by side on the CCD detector. This method ensures that any instrumental shifts affect both spectra equally, minimizing systematic errors in the wavelength calibration.

Figure \ref{fig:Th-Ar} shows a simplified schematic of the method. A stellar spectrum
from the telescope illuminates an optical fiber, whereas a second optical fiber is fed with
Th-Ar lamp. Therefore, both Th-Ar and Stellar spectra are recorded simultaneously, and
we can associate the CCD pixel coordinates with wavelengths given the well-known Th-Ar spectra.

The advantages of using Th-Ar lamps include their well-understood spectrum, thanks to decades of usage in the astronomical community. Th-Ar lamps offer broad wavelength coverage in the optical region from 300-700\,nm, making them ideal for CCD detectors. Th-Ar calibration does not lead to light loss of the stellar spectra, which is beneficial for other types of stellar analysis, such as studying spectral abundance or line shapes.

However, Th-Ar lamps also have some well-known disadvantages. Beyond 650 nm, the density of Th-Ar emission lines decreases rapidly, making these lamps less suitable for calibration in the near-infrared. Another issue is that Th-Ar lamps "age", causing their emission spectra to change over time and "lose" lines, which poses a problem for consistent wavelength calibration over decades. Some Th-Ar lines are so bright that they lead to saturated lines, which "spill" over several spectral orders and add significant line contamination. Finally, aside from astronomy, Th-Ar lamps are not widely used, leading manufacturers to discontinue their production, making such lamps currently scarce.

\subsubsection{Fabry–Pérot etalons and  Laser Frequency Combs}
\label{chap2.2.3}

Traditional calibrators like Th-Ar lamps and iodine cells provide long-term stability and high precision in RV measurements. However, achieving a Doppler precision of 1–10 cm,s$^{-1}$ requires an additional layer of complex optomechanical engineering. In recent years, two advanced calibrators have been employed in combination with standard wavelength calibrators: Fabry–Pérot (F-P) etalons and Laser Frequency Combs (LFCs). The nature and implementation of the F-P and LFC are complex and will not be discussed in detail in this work. For a brief, yet very comprehensive explanation of these methods, we encourage the reader to refer to \citet{Hatzes2019book}.

F-P etalons produce a dense spacing of transmission peaks that can be used for wavelength calibration. When actively controlled using atomic references, they can achieve Doppler precisions of a few cm\,s$^{-1}$. Despite their benefits, F-P etalons pose challenges due to the need for precise calibration of their optomechanical components. Their use, in combination with other calibration methods such as Th-Ar calibrators, can effectively correct instrumental drifts and improve wavelength calibration, thereby enhancing the overall precision of RV measurements.

LFCs provide a dense array of laser peaks, equally spaced in frequency over a broad bandwidth. This makes them superior to traditional calibrators like Th-Ar lamps and iodine cells, as they offer long-term stability, high precision, and known absolute wavelengths (when used together with ThAr lamps). LFCs have shown the potential to achieve RV precisions of about 1 cm\,s$^{-1}$, which is critical for detecting small exoplanets. However, LFCs are extremely complex and costly devices, and their usage is still in the experimental phase.

\begin{figure}[t]
\centering
\includegraphics[width=.456\textwidth]{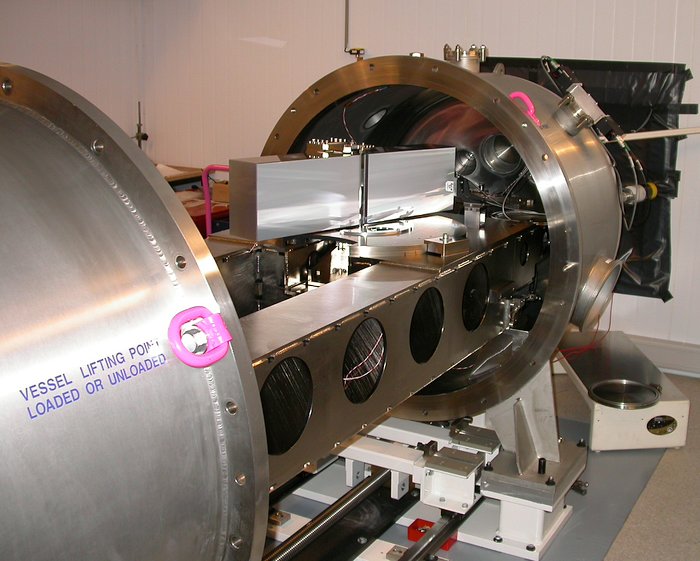}\hspace{0.2cm}
\includegraphics[width=.49\textwidth]{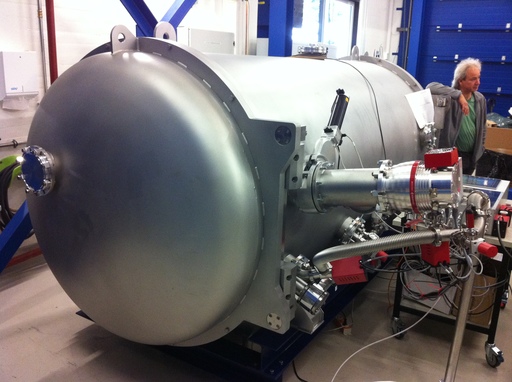}

\caption{The left panel shows the optical bench of the HARPS spectrograph, which resides inside a vacuum tank (opened in the picture) to maintain thermal stability. Credit: ESO. The right panel shows the vacuum tank of the CARMENES optical channel, with the CARMENES system engineer Walter Seifert for scale. Credit: CARMENES Consortium.}
\label{fig_harps}
\end{figure}

\subsection{Mechanical and thermal stability}
\label{chap2.3}

Thermal and mechanical stability are essential for long-term, precise RV measurements. To detect small, rocky exoplanets, the current gold standard employs simultaneous Th-Ar calibrated, ultra-stable \'{E}chelle spectrographs such as HARPS and CARMENES. Figure \ref{fig_harps} depicts the HARPS and CARMENES instruments during their commissioning phase. For instance, HARPS is housed within a specialized vacuum vessel chamber, located in a dedicated room at the 3.6-meter telescope in La Silla, Chile, where access is strictly prohibited. The long-term RV precision of instruments like HARPS is due to their design, which includes robust insulation against temperature and pressure fluctuations and mechanical shocks, maintaining stability for years. This minimizes the thermal effects that can affect the mechanical and optical components of the instrument, thus reducing noise and systematic errors in the wavelength calibration and the point-spread-function (PSF) instrument profile. Currently, the temperature stability of HARPS is maintained below 0.01\,K \citep{LoCurto2024}.

HARPS serves as a benchmark for ultra-precision RV instrumentation. No modern spectrograph designed to detect small rocky exoplanets is built or currently under construction without significant investments in thermal and mechanical stability. As technological advancements continue to extend the limits of achievable RV precision, addressing these instrumental and calibration challenges remains crucial for discovering Earth-like exoplanets in the habitable zones of stars.

\subsection{RV Precision across Spectral Types}
\label{chap2.4}

\begin{figure}[t]
\centering
\includegraphics[width=.995\textwidth]{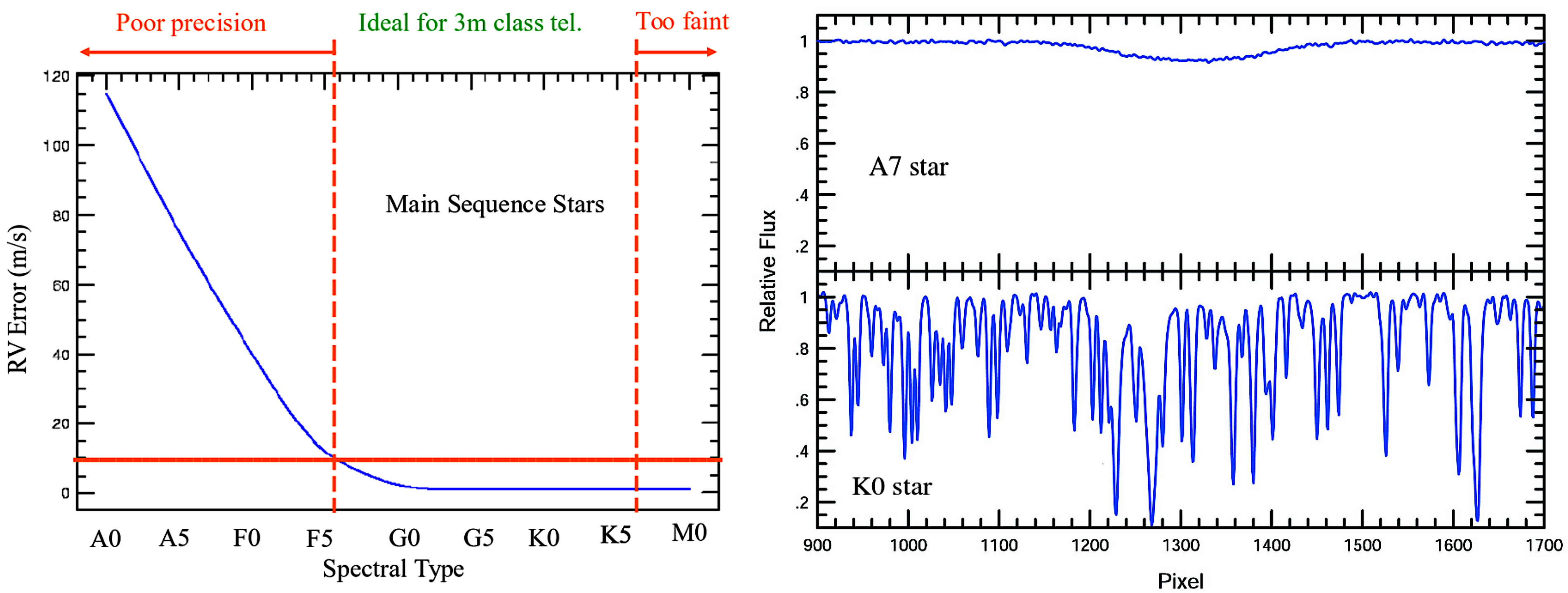}
\caption{The Left panel shows the typical RV precision as a function of stellar spectral type.
The right panel shows the difference in spectra between a young massive star of spectral type A7 and
a star of spectral type K0. The horizontal dashed line marks an RV precision of 10 m\,s$^{-1}$, necessary to
detect a Jupiter-analog around a sun-like star. Credit: both figures are from Artie Hatzes (priv. communication).}
\label{sp_type_vs_rv}
\end{figure}

For a good reason, the RV exoplanet surveys mainly focus on stars with spectral types later than F5. The left panel of Fig. \ref{sp_type_vs_rv} shows the achievable RV precision as a function of stellar spectral type, while the right panel presents an example of recorded spectra for stars with spectral types A7 and K0. Stars earlier than F5 are typically younger, hotter, and rotate faster. As a result, they have fewer and more broadened stellar lines, making it difficult to measure RVs accurately. Therefore, young and hot stars are poor targets for precision RV work.
Cooler stars of spectral type later than F5 have slower stellar rotation velocities. These stars are typically main-sequence (MS) and red giant post-MS stars of spectral types G and K. They possess many deep spectral lines, each containing Doppler information, which allows for the assembly of adequate RV statistics. Consequently, these stars are excellent targets for precision RV surveys. Moreover, most observed red giants of spectral types G and K are retired A-F intermediate-mass stars, for which reliable RV measurements cannot be obtained. Therefore, studying the exoplanet population around red giants provides valuable insights into the population around more massive stars than the Sun.
From Figure \ref{sp_type_vs_rv}, it is evident that these stars are ideal for precision RV measurements using relatively small telescopes, in the 2-3\,meter class, and even smaller telescopes. Small telescopes are much cheaper to operate than larger 8-meter class telescopes, such as the VLT. Consequently, they are often used for RV surveys to take advantage of the lower operational costs while still achieving precise measurements.

As indicated in  Figure \ref{sp_type_vs_rv} on the other side of the spectral classification, stars later than K5 were often too faint for practical observation with small telescopes. Late K and M red-giants are bright and suitable \citep[e.g.,][]{Reffert2015}, but these very evolved stars have been shown to show significant stellar activity, likely due to non-radial stellar pulsations.
Late spectral type MS stars are the most abundant in the solar neighborhood. For instance, M dwarfs represent a large fraction -- about 72\% -- of the stars within 10 pc \citep{Golovin2023}.
While these stars are faint, they have presented significant interest to the scientific community. Exoplanets around K and M-dwarfs induce higher Doppler signals than in more massive stars such as the Sun
because of the smaller mass ratio. The lower stellar masses allow for the detection of potentially rocky planets in the habitable zone (HZ), and multiple-planet systems with relatively packed shorter periods with typical RV semi-amplitudes of a few m\,s$^{-1}$.

For this reason, in recent years, it has become evident that acquiring precise RV measurements in the redder part of the spectrum significantly enhances M-dwarf surveys \citep{Reiners2018b}. Among the instruments tailored for this purpose is the CARMENES spectrograph, which is specifically suited for measurements of M dwarfs. This is due to its operation in the red part of the visible and near-infrared (NIR) spectral regions, which reduces the influence of stellar activity on the RVs.

Consequently, most host stars of RV-detected planets fall within a mass range of 0.4$-$1.2 M$_\odot$, thus presenting a notable bias in the exoplanet demographics as a function of the stellar host's properties.

\noindent
\vspace{10pt}
\rule{\textwidth}{2pt}
\vspace{5pt}

\section{Modeling Radial Velocity curves}
\label{chap3}

\subsection{Extracting Keplerian orbits}
\label{chap3.2}

To introduce the extraction of Keplerian orbitals from RV signals, we start with the simplest
case of an exoplanet orbiting its parent star in a perfectly circular orbit with eccentricity
$e = 0$. In this case, the argument of periastron $\omega$ becomes undefined because the
concept of the point of closest approach to the star is not applicable. It is important to
note that the Doppler method does not allow for the determination of the exoplanet’s orbit in
three dimensions, as the orbital inclination $i$ and the angle giving the orientation of the
orbit on the plane of the sky, called the longitude of the ascending node $\Omega$, remain
undefined. Therefore, for a companion on a perfectly circular orbit the parameters $e$, $\omega$, $i$, and $\Omega$ do not contribute, and the stellar RV around the barycenter can be modeled by a cosine (or sine) model in the form:

\begin{equation}
V(t) = K \cos\left( \frac{2\pi}{P} (t - t_0)\right) + \gamma
    \label{eq:circ}
\end{equation}

\noindent
where $K$ is the semi-amplitude of the RV signal, $P$ is the period of the exoplanet's orbit, $t$ is the time, and $t_0$ is the reference epoch (often the time of periastron passage $t_p$ when $e > 0$). Finally, $\gamma$ in this context is the "absolute" radial velocity of the star with respect to the observer, which can be ignored at this point. In this case, $\frac{2\pi}{P}$ is the mean motion $n$ of the planet, and together with the time after the reference epoch $t - t_0$, it defines the phase of the signal.

Including the eccentricity in the signal, modeling makes it more complex due to the non-sinusoidal, asymmetric, and non-linear nature of the RV curve. This leads to the so-called "RV equation," which for a single planet with an eccentric orbit has the following form:

\begin{equation}
    V(t)=K\left[\cos \left(\nu(t)+\omega\right)+e \cos\omega\right]+\gamma
    \label{eq:radial_velocity_equation}
\end{equation}
where $\omega$ is the argument of periastron of the star (since we measure the stellar RV movement, not the planet's), and we introduce another important time-dependent angle called the true anomaly $\nu(t)$, which is the angle normally used to characterize an observational orbit. To determine the function $\nu(t)$ over time, we must introduce two more important time-dependent angles known as orbital "anomalies": the eccentric anomaly $E(t)$ and the mean anomaly $M_a(t)$. The eccentric anomaly $E(t)$ is an angle used to describe the position of a planet in its elliptical orbit around a star, from which we can derive $\nu(t)$ using the following geometrical relation:

\begin{equation}
\cos E(t) = \frac{\cos \nu(t) + e}{1 + e \cos \nu(t)}
\end{equation}

\noindent
or, alternatively written, $\nu(t)$ can be derived from the $E(t)$ :

\begin{equation}
\tan \left( \frac{\nu(t)}{2} \right) = \sqrt{\frac{1+e}{1-e}} \tan \left( \frac{E(t)}{2} \right)
\end{equation}

\begin{figure}[t]
\centering
\includegraphics[width=.995\textwidth]{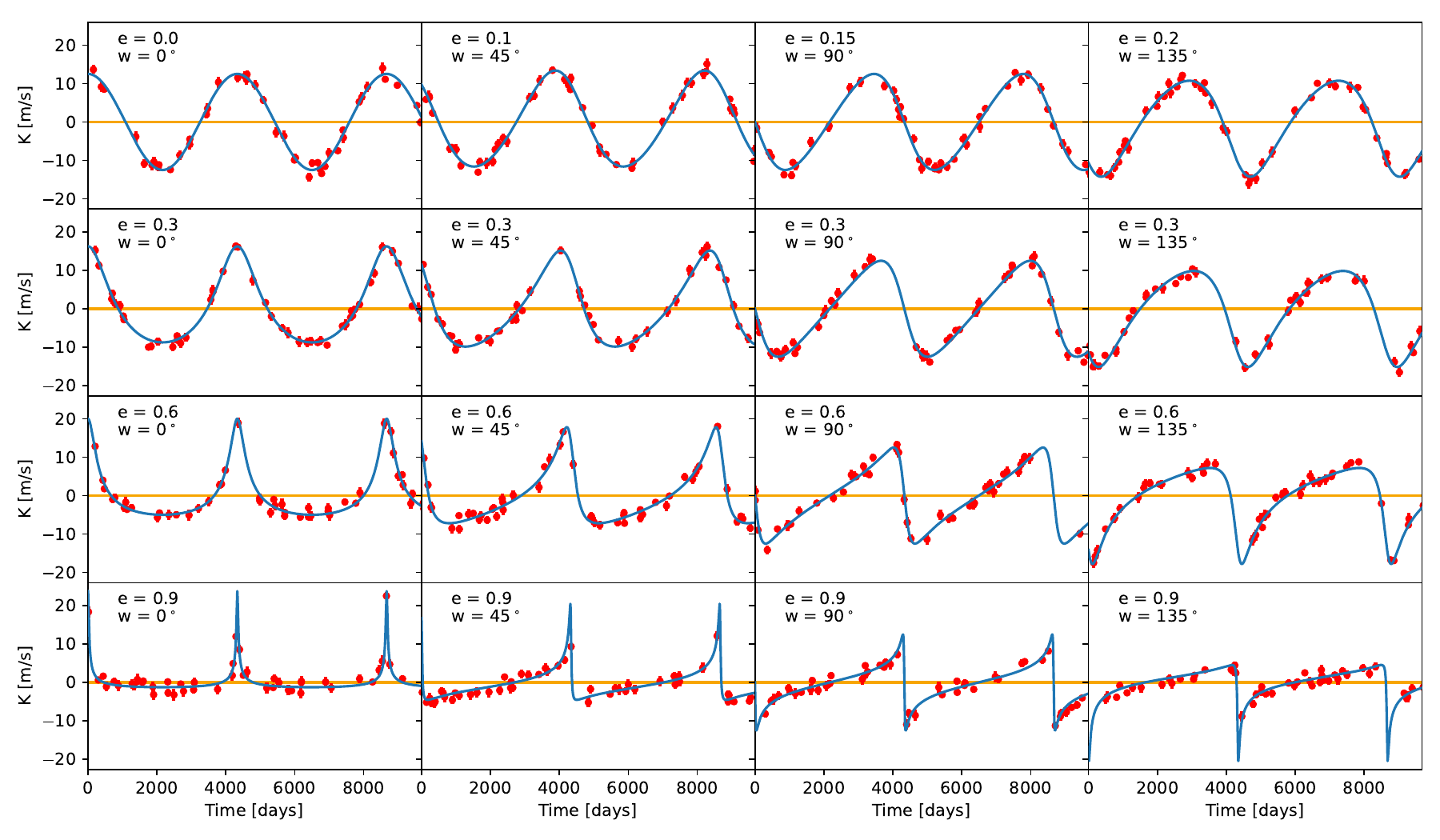}
\caption{RV signature of a single planet, consistent with the period and mass of Jupiter around a 1 $M_\odot$ star for different orbital geometries in terms of eccentricity $e$ and argument of periastron $\omega$. Red data are simulated RVs with the typical precision of HAPRS for solar-type stars.}
\label{Rv_signal_jup_vs_e}
\end{figure}

\noindent
We still have to solve $E(t)$, and for the purpose we need the mean anomaly $M_a(t)$ defined as:

\begin{equation}
    M_a(t) = \frac{2 \pi}{P} (t-t_p) = n(t-t_p)
    \label{eq:mean_anomaly}
\end{equation}

\noindent
which tells us how far an exoplanet has progressed with respect to the time of periastron passage $t_p$, or other arbitrary epoch.
The eccentric anomaly $E(t)$ is related to the mean anomaly $M(t)$ through Kepler's Equation:

\begin{equation}
M_a(t) = E(t) - e \sin E(t)
\end{equation}

\noindent
which for a perfectly circular orbit with $e=0$ it leads to Eq.\ref{eq:circ}. Unfortunately, Kepler's equation cannot be solved analytically and needs some
iterative approach. Fortunately, however, modern computational algorithms can solve this equation numerically with relative ease.

At this point, we are ready to model the RV shape of an exoplanet signal, and the only parameter that requires attention is the RV semi-amplitude $K$. In signal modeling, $K$
is a fitting parameter independent of the physical reason for the observed semi-amplitude. Yet, we know that $K$ provides information about the star-planet mass ratio, and if we
know the mass of the star, from e.g., stellar isochrone models, which is often the case, we can get the minimum mass of the planet. The radial velocity semi-amplitude is expressed as:

\begin{equation}
    K = \frac{2\pi}{P} \frac{a \sin i}{\sqrt{1 - e^2}}.
\end{equation}

\noindent
or alternatively one can get the projected semi-major axis of the star, measuring the RV signature of $K$, $P$, and $e$:

\begin{equation}
    a \sin i = \frac{K P \sqrt{1 - e^2}}{2\pi}.
\end{equation}

\noindent
The value of the projected semi-major axis $a \sin i$, however, is a problem since $a$ nor $\sin i$ can be determined separately. After making some transformations using Kepler's third law, the orbit of the star around the barycenter is given by the form:

\begin{equation}
P = \sqrt{\frac{4\pi^2}{GM}a^3}
\end{equation}

\noindent
where G is the gravitational constant and the total mass M, which to account for the unknown $\sin i$ leads to the mass function:
\begin{equation}
f(m) = \frac{m_p^3 \sin^3 i}{(M_\star + m_p)^2}.
\end{equation}

\noindent
and after some more transformations, we end up with the most common expression of the RV semi-amplitude:

\begin{equation}
    K = \left(\frac{2\pi G}{P}\right)^{1/3} \frac{m_p \sin i}{(M_\star + m_p)^{2/3}} \frac{1}{\sqrt{1 - e^2}}.
\end{equation}

\noindent
Estimates of $M_\star$ can be inferred from stellar models, which, rely on the observed photometric data, stellar luminosity, and effective temperature. From RV data fitting the spectroscopic observables $K$, $P$, and $e$ could be used to estimate 
the minimum mass of the planet $m_p \sin i$, which is of our primary interest. Assuming the mass of the planet is $ m_p \ll M_\star$, thus approximating to $m_p + M_\star \approx M_\star$, A common practical equation for K is often given as:

\begin{equation}
K = \frac{28.4329\text{m\,s}^{-1}}{\sqrt{1 - e^2}} \,  \left( \frac{P}{1 \, \text{yr}} \right)^{-\frac{1}{3}} \left( \frac{m_p \sin i}{M_{\rm Jup}} \right) \left( \frac{M_\star}{M_{\odot}} \right)^{-\frac{2}{3}}
\end{equation}

Table \ref{table_K} summarizes the period, eccentricity, semi-major axes, and calculated semi-amplitudes $K$ of the planets in the Solar system. It is clear that if an alien civilization observes the Sun as a star and obtains stellar RVs with similar precision to what we currently have on Earth, the only planets likely to be detected are Jupiter and Saturn, with $K = 12.5$ m\,s$^{-1}$ and $2.8$ m\,s$^{-1}$, respectively.

\begin{table*}
  \centering
  \caption{Orbital parameters and semi-amplitude of the planets in the Solar system.}
\begin{tabular}{p{3.0cm} p{3.0cm} p{3.0cm}  p{3.0cm} p{2.0cm}}

\hline
Planet & P (years) & $e$ & a (au) &  $K$ (m\,s$^{-1}$) \\
\hline
Mercury & 0.24 & 0.2056 & 0.39 & 0.008 \\
Venus & 0.62 & 0.0067 & 0.72 & 0.086 \\
Earth & 1.00 & 0.0167 & 1.00 & 0.089 \\
Mars & 1.88 & 0.0934 & 1.52 & 0.008 \\
Jupiter & 11.86 & 0.0489 & 5.20 & 12.48 \\
Saturn & 29.46 & 0.0565 & 9.55 & 2.76 \\
Uranus & 84.01 & 0.0463 & 19.22 & 0.30 \\
Neptune & 164.79 & 0.0097 & 30.11 & 0.28 \\

\hline
\end{tabular}
\label{table_K}
\end{table*}

Figure \ref{Rv_signal_jup_vs_e} shows the typical RV signature of simulated RVs induced by an exoplanet with $1\,M_{\rm jup.}$ at 5.2 au, orbiting a star with $1\,M_\odot$ (a Jupiter analog) for different orbital eccentricities and geometries in terms of $\omega$. The simplest case is for $e=0, \omega=0$, shown in the top left panel, which is practically a simple sine model. By increasing $e$ and varying $\omega$, the signal becomes more complex.

Now we have a model with parameters $V(\theta)$, which are the five spectroscopic observables related to the star’s Keplerian orbit: $\theta = \{e, P, M_a, \omega, K\}$ and the systemic RV velocity $\gamma$. If more than one planet is evident in the data, then a second set of these parameters should be included in the model, i.e., $\theta_{n} = \{e_n, P_n, M_{a,n}, \omega_n, K_n\}$, where $n$ is the number of planets in the model. If more than one RV dataset is used, which is often the case (see Figs. \ref{fig3} and \ref{fig4}), then we must also fit for the RV data offset $\gamma_{p}$ for each dataset, where $p$ is the number of datasets. Then we can find the  maximum of the $\ln\mathcal{L}$ function:

\begin{equation}
\ln\mathcal{L}(\theta) = -\frac{1}{2} \sum_{i=1}^{N} \left[ \frac{(v_i - V(t_i; \theta))^2}{\sigma_i^2} + \ln\left(\sqrt{2\pi} \sigma_i\right) \right]
\label{Lnl}
\end{equation}

\noindent
where $v_{i}$ are the observed radial velocities, $V(t_i; \theta)$ are the model radial velocities at time $t_i$, and $\sigma_i$ are the uncertainties in the observed radial velocities.

The $\ln\mathcal{L}(\theta)$ function can be optimized using standard numerical techniques such as the Nelder-Mead simplex algorithm \citep[][]{NelderMead}, which converges to the best-fit solution, or through parameter sampling techniques such as Markov chain Monte Carlo (MCMC) sampling and the
nested sampling methods, which are used for Bayesian inference analysis and posterior estimation of the sampled parameters consistent with the data.

\subsection{Planet search using periodogram analysis}
\label{chap3.1}

Detecting exoplanets using RV data cannot be done by eye, particularly for short-period planets with sufficient temporal baselines of observations. An additional challenge in the preliminary exoplanet vetting process is the complexity of RV data, which is profoundly affected by multiple RV signals (i.e., multiple-planet systems), significant RV noise (see Sec.\ref{chap4}), uneven time sampling, and observational gaps due to Earth's rotation and seasonal changes. These factors can produce suspicious signals due to aliasing, making the RV data analysis ambiguous and complex.

Astronomers rely on automatic period search algorithms based on Fourier data analysis principles. Fourier's work demonstrated that any continuous function could be represented by a series of sines and cosines, simplifying the detection of periodic signals in time-series data. In the domain of RV data, this translates to identifying peaks in the Fourier transform that correspond to the frequencies of orbiting exoplanets. A popular tool for astronomers is the so-called ``generalized'' Lomb-Scargle periodogram \citep[GLS][]{ Zechmeister2009}. The GLS is very efficient and fast at revealing dominant frequencies despite the data's irregular sampling and gaps, offering a robust way to analyze noisy datasets. In recent years, even more complex and accurate, yet CPU-intensive derivatives of the GLS have gained popularity. These include the maximum $\ln\mathcal{L}$ periodogram (MLP), which is more robust for multi-telescope data, the Bayesian generalized Lomb-Scargle periodogram (BGLS), and others.

Periodograms provide the numerical stability and speed essential for managing the large volumes of data typically involved in exoplanet detection. Hence, periodograms remain a cornerstone technique for astronomers and astrophysicists engaged in the search for exoplanets through RV measurements, providing a critical tool for interpreting complex astronomical RV data.

\subsection{Exoplanet characterization tools}
\label{chap3.2}

\begin{figure}[t]
\centering
\includegraphics[width=.995\textwidth]{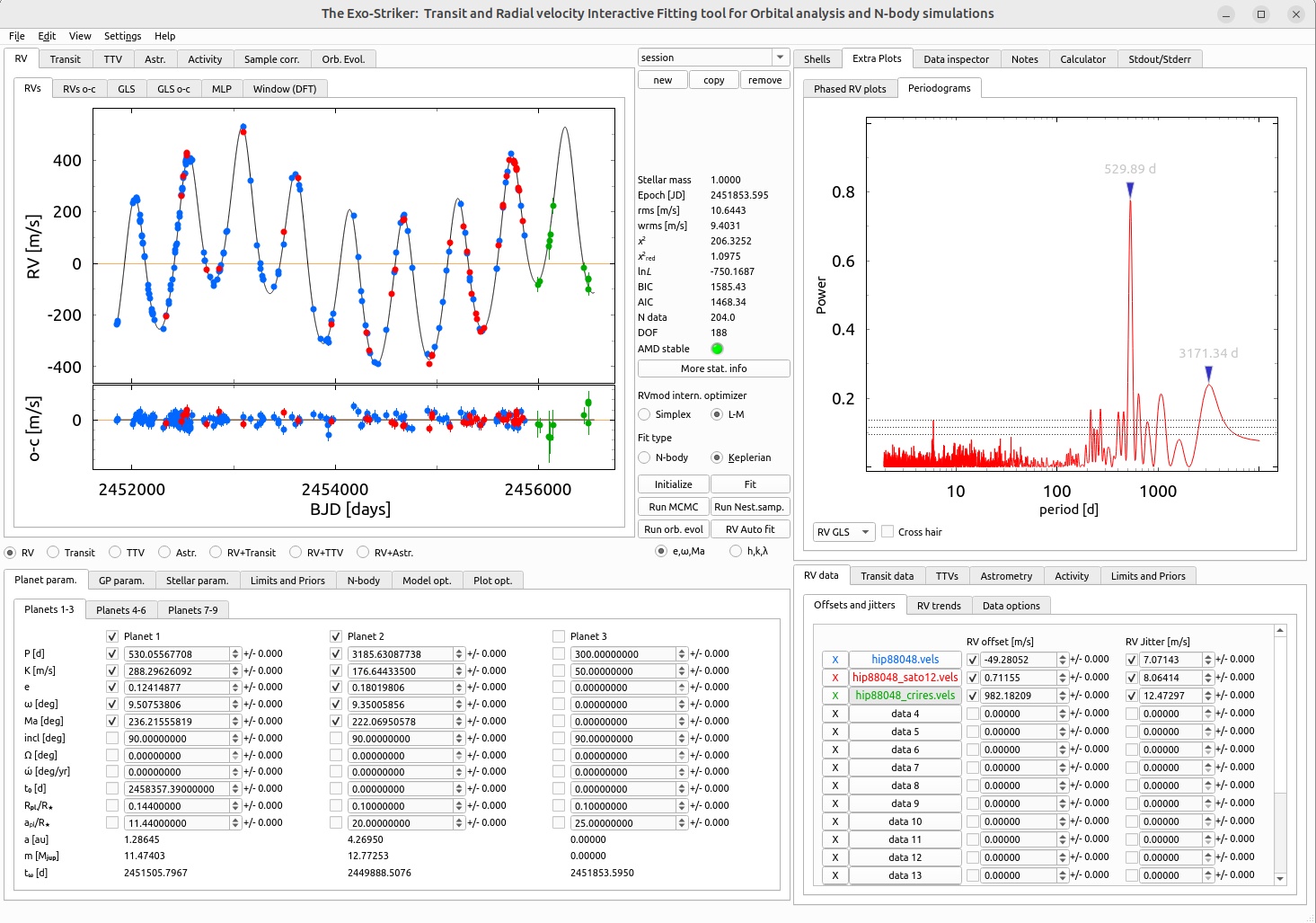}
\caption{GUI screenshot of the {\sc Exo-Striker} tool analyzing the $\nu$ Oph multiple-planet system. The left panel shows a typical view of the GLS and RV fitting part, which provides fast plotting, parameter overview, and statistics.
}
\label{ES_tool}
\end{figure}

Several exoplanet modeling programs are publicly available, with most developed in the modern {\sc Python3} language, designed to analyze exoplanetary orbits from data such as precise RVs and transit photometry.
For instance, the  {\sc Exo-Striker} exoplanet toolbox could be of particular interest to the reader. This is an open-source program with a user-friendly Graphical User Interface (GUI), available for download at \url{https://github.com/3fon3fonov/exostriker}.
Figure \ref{ES_tool} shows a typical GUI view of the {\sc Exo-Striker} exoplanet toolbox. Developed for both professional and educational use, {\sc Exo-Striker} is suitable for postgraduate students and their tutors. It offers a comprehensive suite of features for the detailed analysis of both transit and RV data. This includes power spectrum analysis such as the GLS periodogram for identifying signals, Keplerian and dynamical modeling of multi-planet systems, and standard frequentist and Bayesian methods for orbital parameter estimation. The {\sc Exo-Striker} also provides tools for long-term stability checks of multi-planet systems and generating fast interactive plots. Compatible with MacOS, Linux, and Windows, {\sc Exo-Striker} is a versatile choice for astronomers and astrophysicists engaged in exoplanet discovery and characterization.

The orbital fits for HR\,6388 B, 51 Peg\,b, and Proxima Cen\,b discussed in this work were conducted using the {\sc Exo-Striker}, utilizing publicly available literature and archival RV data for these targets. The {\sc Exo-Striker} sessions for these analyses are available here \url{https://zenodo.org/records/13285406}.

\noindent
\vspace{10pt}
\rule{\textwidth}{2pt}
\vspace{5pt}

\section{Stellar activity and false positive detections}
\label{chap4}

Achieving m\,s$^{-1}$ and even cm\,s$^{-1}$  precision in RV measurements is a complex
task due to various instrumental and observational limitations, but also
due to Stellar activity. The influence of stellar activity on RV measurements is complex and perhaps the most crucial topic in the search for exoplanets. The implications of stellar activity are numerous and cannot be fully discussed in this work. Therefore, for a more detailed yet easy-to-understand overview of the different stellar activity-induced effects on RV data, we recommend Chapters 9 and 10 in \citep{Hatzes2019book}.

Here, we discuss the most prominent astrophysical limitations, which, separately or combined, are the main reason for hampering the firm detection of low-mass exoplanets similar to that of Earth.


\subsection{Stellar jitter}
\label{chap4.2}

A significant source of RV noise originates from astrophysical phenomena within stars, over which astronomers have literally no control. Several intrinsic stellar phenomena contribute to what is known as ''RV jitter'', with solar-type p-mode oscillations being the dominant process. In the Sun, these p-mode pulsations are attributed to the periodic expansion and contraction of the solar surface, producing small, correlated RV noise with time scales of approximately 5 minutes. Given the typically sparse cadence of exoplanet surveys, this correlated noise appears as stochastic RV noise. Every star exhibits an ''RV jitter'' and its amplitude and timescale depend on various factors such as stellar type, age, and radius. For instance, G-type solar stars appear relatively quiet during the MS phase but show a significant increase in jitter as they evolve onto the red giant branch \citep{Kjeldsen2011}.

RV jitter presents a major challenge in exoplanet research, particularly when detecting and characterizing very small exoplanets. It is crucial to account for RV jitter when analyzing radial velocity data to ensure accurate interpretation of any detected signals that could potentially originate from exoplanets. One strategy to mitigate the effects of jitter is to accumulate a large number of RV observations, thereby enhancing the significance of exoplanet signals relative to the RV noise. Although this approach is costly, it is perhaps the only feasible method to detect small Earth-mass planets around solar-type stars. Fortunately, astronomers have developed effective numerical techniques to manage RV jitter. When fitting orbits and estimating uncertainties in orbital parameters, we can include the RV jitter term in quadrature to the total error budget, treating RV jitter as originating from stochastic processes:

\begin{equation}
\sigma_{i,n} = \sqrt{\sigma^2_{i} + \sigma^2_{\rm jitter,n}}
\end{equation}

\noindent
Here, for each RV dataset $n$, we add the RV jitter term $\sigma_{\rm jitter,n}$ to each $i$-th uncertainty. It is important to understand that what we call "RV jitter" may, in fact, result from a combination of astrophysical and instrumental noise. Therefore, we adopt different jitter terms for each RV dataset.

Following the prescription of \citet{Baluev2009} of incorporating a parameter term for the stochastic stellar RV jitter for each used RV dataset into the $\ln\mathcal{L}$ function, the log-likelihood function now has the form:

\begin{equation}
\ln\mathcal{L}(\theta) = -\frac{1}{2} \sum_{i=1}^{N} \left[ \frac{(v_i - V(t_i; \theta))^2}{\sigma_i^2 + \sigma_{\rm jitter,n}^2} + \ln\left(\sqrt{2\pi (\sigma_i^2 + \sigma_{\rm jitter,n}^2)}\right) \right]
\end{equation}

This inclusion helps obtain a realistic estimation of the RV scatter and ensures the integrity of the derived orbital solutions and estimated parameter uncertainties. High-resolution spectrographs and advanced data processing techniques are crucial in dissecting these contributions to jitter. Ultimately, understanding and correcting for RV jitter is essential for the reliable detection of Earth-like exoplanets.

\subsection{Stellar Activity: A Source of False Exoplanet Detections}
\label{chap4.2.1}

One significant source of astrophysical noise is induced by stellar rotational modulation, which can severely impact exoplanet detectability. Stars typically rotate over a period of days, and stars similar to the Sun have spots and plages that move across the stellar disk as the star rotates. This phenomenon causes flux variations in the stellar regions they traverse, leading to differences in weight of the blue-shifted and red-shifted part of the stellar disk. These shifts create stellar line asymmetries, which our RV extraction pipelines interpret as Doppler shifts. Since these signals could be periodic, they can mimic the RV signals from exoplanets.

The challenge of distinguishing between signals caused by stellar activity and those due to gravitational influences from exoplanets is significant. Stellar activity can introduce RV variations that are often indistinguishable from those caused by orbiting planets without careful analysis. This complexity is compounded by the effects of the rotation period $P_{\text{rot}}$ and its harmonics $\left(P_{\text{rot}}/2, P_{\text{rot}}/3, \ldots\right)$, which can appear in the RV data depending on the number and arrangement of active regions on the stellar surface. Accurately identifying these signals requires sophisticated techniques like Fourier analysis to parse out the fundamental rotational frequency from its harmonics.

Stellar magnetic activity also plays a crucial role, particularly through phenomena like sunspots, faculae, and flares, which cause localized surface velocity and brightness changes, leading to further complications in RV data analysis. For instance, the $\sim$11-year Solar cycles can induce RV variations with amplitudes and periods similar to those of Jupiter.

Stellar granulation is another significant source of stellar RV jitter, arising from the convective motion within a star's photosphere. This granulation leads to varying RV signals as different parts of the star's surface move relative to each other, contributing to the overall noise level in RV measurements.

Stellar activity and its effects on RV measurements are more pronounced in young stars. These stars, particularly those younger than 1 G\,yr, exhibit small rotational periods that can closely align with the orbital periods of potential exoplanets, complicating the differentiation between stellar and planetary signals. The magnetic dynamo, which drives this activity, is stronger in younger stars, leading to larger and more numerous active regions. As stars age, their rotation slows, and the dynamo weakens, reducing the size and impact of active regions.

For exoplanet searches, this necessitates focusing on older, slowly rotating stars where activity levels are lower and less likely to interfere with RV measurements. On the other hand, extreme cases such as evolved old stars such as K-giants are suspected to exhibit non-radial pulsations, which can also mimic planets \citep{Hatzes2018}.
Techniques such as observing in the near-infrared domain have been proposed to mitigate these issues, as the contrast of dark spots is reduced at these wavelengths. Doppler signals due to orbiting companions are wavelength-independent. Additionally, the use of spectroscopic indicators like the line bisector \citep{Queloz2001} and the Ca II H\&K chromosphere index help in identifying and correcting for activity-related distortions in RV data.

Achieving high precision in RV measurements and confidently confirming the presence of exoplanets amidst the noise of stellar activity is an ongoing challenge. One effective strategy is to densely sample and average data over periods comparable to the star's rotation period. This approach helps enhance precision and manage the influence of stellar activity. By understanding and accounting for these factors, researchers can improve the accuracy of exoplanet detection and avoid mistakenly interpreting stellar noise as planetary signals.

\noindent
\vspace{10pt}
\rule{\textwidth}{2pt}
\vspace{5pt}

\section{The future}
\label{chap5}

Technological advancements in spectroscopy aim to develop highly stable, high-resolution spectrographs capable of measuring RVs down to 10 cm\,s$^{-1}$, in the near future.
The primary goal of reaching such extreme RV precision is to enable the identification of Earth-like exoplanets in the HZ around stars similar to our Sun.
Notable projects in this ambitious field include "The Terra Hunting Experiment" \citep[][]{Thompson2016}, the EXtreme PREcision Spectrometer \citep[EXPRES,][]{Jurgenson2016}, and the "Second Earth Initiative Spectrograph (2ES)" (Buchhave et al. in prep.).
These future Doppler surveys are designed to discover Earth-mass exoplanets with orbital periods close to one year around nearby Sun-like stars. The \'{E}chelle instruments being developed for these surveys will utilize significantly improved versions of existing instruments like the
HARPS and MAROON-X spectrographs, which are already capable of reaching precision well below 1 m\,s$^{-1}$, will be installed on 3-meter class telescopes. These surveys aim to collect numerous  RV measurements in an automated fashion for a relatively small sample of nearby Sun-like stars over a decade-long period.

However, achieving a precision of 10 cm\,s$^{-1}$ is only the first major milestone. A deeper understanding of stellar activity is needed, and the exoplanet field is actively working in this direction. Fortunately, large-scale insights into stellar activity are recorded in the spectra, and future high-resolution spectrographs will certainly help tackle the challenge of stellar noise. By continuously monitoring stars over many years on a high cadence, researchers aim to overcome instrumental and stellar activity signals, providing unprecedented insights into the presence of terrestrial planets in our solar neighborhood.

Another important aspect of RV detection is the continuation of RV surveys, which, by increasing the temporal baseline of the RVs, will reveal the population of cold Jovian analogs located at orbital distances between 5-20\,au, similar to the giants in our Solar System. This exoplanet demographics region is still to be explored (see Fig.\ref{fig3}).

In terms of high-resolution, high-SNR spectra, some of these spectrographs, which are planned to be utilized on 8-meter class telescopes or future extremely large telescopes (ELTs), would be ideal for exoplanet atmosphere characterization of rocky planets using transmission spectroscopy. Apart from the ongoing efforts in recording ultra-precise RV measurements and transmission and emission spectroscopy using ESPRESSO \citep{Pepe2021} at the VLT in Chile, another notable project must be mentioned. The ArmazoNes high Dispersion \'{E}chelle Spectrograph (ANDES) is a high-resolution optical and near-infrared spectrograph being developed for the Extremely Large Telescope (ELT). ANDES is expected to revolutionize exoplanet science by enabling the detection of bio-signatures in exoplanet atmospheres and providing detailed observations of planetary systems \citep{Palle2023}.

The increased RV sensitivity of future spectrographs, combined with existing ones, will enhance the integration of the RV method with other exoplanet detection techniques, such as transit photometry, direct imaging, and forthcoming data from Gaia DR4. The RV method will continue to be vital for characterizing exoplanets, whether used alone or in combination with other techniques.

The potential deployment of space-based RV observatories could revolutionize the method, though currently, this idea is seen as too expensive and technically impractical. The absence of atmospheric disturbances in space would enhance data accuracy and sensitivity. However, deploying high-resolution, high-precision \'{E}chelle spectrographs and managing their complex data collection from space presents significant challenges and risks.

Ground-based high-resolution spectroscopy will continue to play a crucial role in characterizing exoplanet atmospheres. As detection capabilities improve, so will the ability to characterize the atmospheres of smaller, Earth-like planets. This progress will be enhanced by combining RV data with that from direct imaging missions, facilitating detailed studies of the climates and chemical compositions of exoplanet atmospheres.

Over the next thirty years, the RV Doppler method will likely enhance its existing capabilities and expand into new realms of astronomical inquiry.
Our understanding of the Universe and our place within it will deepen, marking an exciting era of exoplanet discoveries and the characterization of their physical properties.

\noindent
\vspace{10pt}
\rule{\textwidth}{2pt}
\vspace{5pt}

\section{Conclusion}
\label{chap6}

The stellar RV Doppler method has remained an invaluable asset in the arena of exoplanet discovery and research. It has made a giant leap in understanding the universe other than our solar system. RV thus remains a cornerstone technique in the detection of exoplanets and precise measurements of their masses.
Constant improvement in spectrographic technologies and enhancement in data analytic methodologies is increasing both sensitivity and accuracy in RV measurements.

In that sense, the integration of the RV data with observation data acquired from other methods, such as transit and astrometric stellar measurements, can fill in missing information about planetary compositions and orbits and give us a detailed and complete view of planetary systems. However, challenges remain, particularly in detecting Earth-like planets due to the subtle signals they produce. The evolution of high-precision spectrographs and advanced statistical techniques has been crucial in overcoming these hurdles and enhancing the sensitivity and accuracy of radial velocity measurements. These technological and methodological advancements are driving a new era in exoplanet discovery, where the detection of Earth analogs around Sun-like stars becomes increasingly feasible, promising profound implications for our understanding of planetary systems across the galaxy. As such, Doppler surveys continue to play a critical role in the field of exoplanet research, shaping future explorations and studies aimed at uncovering the diversity of planets in our Galaxy.

In that sense, the integration of the RV data with observation data from other methods such as
transit and astrometric stellar measurements can complete observations about planetary compositions and orbits, make them complementary, and give us a detailed view of planetary systems. The RV method is bound to bring even bigger discoveries in the future, with improved instruments and techniques, including the detection of Earth twins at Sun-like stars. The RV method, therefore, remains a fundamental tool of exoplanet science, one needed both in the relentless quest for new worlds and in their later detailed characterization.

\begin{ack}[Acknowledgments]

I would like to express my heartfelt gratitude to the following individuals and organizations for their invaluable contributions to this work:
Deyan Mihailov for his assistance in creating some of the figures.
Elena Vchkova for her diligent efforts in extracting the archival RVs for HR\,6388. Adrian Kaminski for the insightful discussions that greatly enriched this work. Artie Hatzes, Martin K\"urster, and Sabine Reffert for reading the manuscript and providing valuable comments.
This work is supported by the Bulgarian National Science Foundation (BNSF) program "VIHREN-2021" project No. KP-06-DV/5.
\end{ack}

\seealso{Many excellent reviews and textbooks are available on the Doppler method for detecting exoplanets. However, I highly recommend \textit{The Doppler Method for the Detection of Exoplanets} by Artie Hatzes \citep{Hatzes2019book}, which served as my primary guide reference while writing this chapter. }

\bibliographystyle{Harvard}
\bibliography{reference}

\end{document}